# Surface plasmon on the metal nanofilament excitation by a quantum oscillator

V.V.Lidsky

The P.N.Lebedev Physical Institute of RAS
119991 Moscow, Russia
vlidsky@sci.lebedev.ru

Surface waves on a thin metal filament are described in terms of quantum electrodynamics. The interaction of surface waves with a quantum oscillator is discussed in the dipole approximation. The increase in the spontaneous emission rate of the excited quantum oscillator, the so called Purcell factor, is evaluated to be as high as ten to the five times.

# Возбуждение поверхностных плазмонов на металлическом наноцилиндре квантовым излучателем

В.В.Лидский

Физический институт им. П. Н. Лебедева РАН
119991 Москва, Ленинский пр., 53
vlidsky@sci.lebedev.ru

А Н Н О Т А Ц И Я

Рассмотрены поверхностные волны на тонком металлическом цилиндре в рамках формализма квантовой электродинамики. Рассмотрено в дипольном приближении взаимодействие поверхностных плазмонов с квантовым излучателем, расположенным в непосредственной близости от поверхности цилиндра. Показано, что вероятность спонтанного перехода с излучением плазмона может на пять порядков превосходить вероятность дипольного перехода в свободном пространстве.

1. После появления в печати сообщений о запуске нанолазеров или SPAZER'ов /1,2/, использующих механизм возбуждения поверхностных волн молекулами красителя, возник вопрос о построении последовательной квантовой теории взаимодействия поверхностных волн с квантовым излучателем. Дело в том, что вероятность спонтанного перехода квантовой системы существенным образом зависит от структуры пространства, окружающего излучатель. Перселл заметил, что вероятность спонтанного перехода возрастает на несколько порядков, если вблизи излучателя находится микроскопическая частица /3/. В /4/ эффект Перселла был привлечен для объяснения наблюдаемого увеличения на 14 порядков сечения комбинационного рассеяния — явления гигантского комбинационного рассеяния (SERS) /5,6/. Теория SPAZER'а была предложена в недавно появившейся работе /7/, где показано, что и здесь учет эффекта Перселла позволяет в принципе преодолеть трудность, вызванную исключительно сильным затуханием поверхностных волн.

В данной работе мы будем следовать методу расчета вероятности, использованному нами в /8/. Мы рассмотрим поверхностные волны, распространяющиеся по тонкому металлическому стержню субволнового диаметра, окруженному диэлектрической средой. Затем мы разложим поле на элементарные моды и определим операторы вторичного квантования, с помощью которых поле поверхностной волны может быть описано с точки зрения квантовой электродинамики. Затем рассмотрим квантовый осциллятор, помещенный вблизи стержня и вычислим вероятность излучения этим осциллятором поверхностного плазмона. Полученную величину сравним с известной формулой для вероятности спонтанного излучения фотона в свободном пространстве и придем к формулировке эффекта Перселла, применительно к излучателю вблизи субволнового цилиндра.

2. Рассмотрим металлический цилиндр малого радиуса $\rho$ в диэлектрической среде (рис. 1). Диэлектрическую проницаемость среды будем считать равной $\varepsilon_h$. Диэлектрическую проницаемость металла будем считать связанной с частотой формулой Друдэ: $\varepsilon_m = 1 - \omega_{pl}^2/\omega^2$. Мнимая часть диэлектрической проницаемости в данной работе учитываться не будет. Нас будет интересовать диапазон частот не превышающих плазменную частоту, таких что соответствующая длина волны в диэлектрике много больше радиуса цилиндра: $2\pi c/\omega \gg \rho$. Магнитную проницаемость металла цилиндра будем считать равной 1.

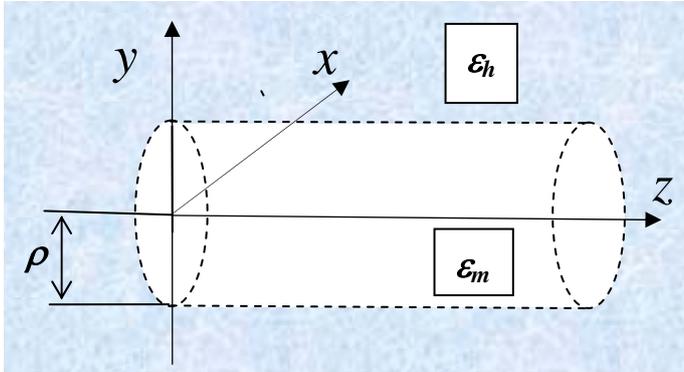

На первом шаге наша задача определить собственные поверхностные волны такого пространства. Уравнения Максвелла при отсутствии внешних токов запишем в виде:

$$rot\ \vec{H} = \varepsilon \cdot \partial_t \vec{E} \qquad rot\ \vec{E} = -\partial_t \vec{H} \qquad (2.1)$$

$$div\ \vec{H} = 0 \qquad div\ \vec{E} = 0 \qquad (2.2)$$

Мы видим, что уравнения (2.2) выполнены, если выполнены (2.1).

Аналогично тому, как это делается в свободном пространстве /9/, выберем конечный участок цилиндра длиной **L**. Разложим поля в ряды Фурье по координатам $z$ и $\varphi$ цилиндрической системы, ось $z$ которой совпадает с осью цилиндра.

$$\vec{E} = \sum_{k=-\infty}^{\infty}\sum_{n=-\infty}^{\infty} \vec{E}_{nk}(t,r) \cdot \exp(ih_k \cdot z + in \cdot \varphi) \qquad \vec{H} = \sum_{k=-\infty}^{\infty}\sum_{n=-\infty}^{\infty} \vec{H}_{nk}(t,r) \cdot \exp(ih_k \cdot z + in \cdot \varphi) \qquad (2.3)$$

Зависимость от времени решений уравнений (2.1) будем искать в виде:
$$\vec{E}_{nk}(t,r), \vec{H}_{nk}(t,r) \propto e^{-i\omega t} \qquad (2.4)$$

причем частота $\omega$ определяется из известного характеристического уравнения, которое будет выписано ниже (см.(3.9-10)). Общий вид решения получим суммируя решения для каждого корня характеристического уравнения. В этих предположениях уравнения Максвелла (2.1) приводят к уравнениям для компонент $\vec{E}_{nk}(t,r), \vec{H}_{nk}(t,r)$:

$$\frac{in}{r}H_z - ihr \cdot H_\varphi = -i\omega\varepsilon \cdot E_r \qquad (2.5)$$

$$\frac{ih}{r}H_r - \frac{1}{r}\partial_r H_z = -i\omega\varepsilon \cdot E_\varphi \qquad (2.6)$$

$$\frac{1}{r}\partial_r(r^2 H_\varphi) - \frac{in}{r} \cdot H_r = -i\omega\varepsilon \cdot E_z \qquad (2.7)$$

$$\frac{in}{r}E_z - ihr \cdot E_\varphi = i\omega \cdot H_r \qquad (2.8)$$

$$\frac{ih}{r}E_r - \frac{1}{r}\partial_r E_z = i\omega \cdot H_\varphi \qquad (2.9)$$

$$\frac{1}{r}\partial_r(r^2 E_\varphi) - \frac{in}{r} \cdot E_r = i\omega \cdot H_z \qquad (2.10)$$

Для удобства записи мы опустили у компонент напряженностей индексы $nk$, указывающие по каким переменным сделано преобразование Фурье. Волновое уравнение приводит к следующему соотношению для спектральных компонент:

$$\left(\partial_r^2 + \frac{1}{r}\partial_r - \frac{n^2}{r^2}\right)E_z = \left(h^2 - \omega^2\varepsilon\right)\cdot E_z \tag{2.11}$$

И аналогичному уравнению для $H_z$.

3. Решение этого уравнения качественно зависит от знака величины $h^2 - \omega^2\varepsilon$. Внутри цилиндра $\varepsilon_m < 0$, следовательно регулярным при $r = 0$ решением уравнения (2.11) окажется модифицированная функция Бесселя:

$$E_z = const \cdot I_n(pr) \tag{3.1}$$

где

$$p = \sqrt{h^2 - \varepsilon_m \cdot \omega^2} \tag{3.2}$$

Вне диэлектрика возможны два варианта. При малых значениях волнового числа $h$ имеет место соотношение $h^2 - \omega^2\varepsilon_h < 0$ и решением (2.11) окажутся осциллирующие цилиндрические функции $J_n(qr), Y_n(qr)$. Эти решения описывают волны, уносящие энергию в пространство. В данной работе мы их рассматривать не будем. Наше исследование посвящено процессам излучения поверхностных волн, имеющих большие волновые числа, так что в дальнейшем будем предполагать $h^2 - \omega^2\varepsilon_h > 0$, а, следовательно, величину

$$q = \sqrt{h^2 - \varepsilon_h \cdot \omega^2} \tag{3.3}$$

будем предполагать вещественной. Убывающим на бесконечности решением уравнения (2.11) оказывается функция Макдональда:

$$E_z = const \cdot K_n(qr) \tag{3.4}$$

Таким образом, мы приходим к решениям (2.11) в металле и диэлектрике:

$$\begin{aligned}
E_z^{(m)}(r,t) &= C_1 \cdot e^{-\omega t} \cdot I_n(pr)/I_n(p\rho) \\
E_z^{(h)}(r,t) &= C_2 \cdot e^{-\omega t} \cdot K_n(qr)/K_n(q\rho) \\
H_z^{(m)}(r,t) &= C_3 \cdot e^{-\omega t} \cdot I_n(pr)/I_n(p\rho) \\
H_z^{(h)}(r,t) &= C_4 \cdot e^{-\omega t} \cdot K_n(qr)/K_n(q\rho)
\end{aligned} \tag{3.5}$$

Из (3.5) и (2.5)-(2.10) легко находятся компоненты $E_\varphi, H_\varphi$ (см. Приложение А):

$$\begin{aligned}
H_\varphi^{(m)}(r,t) &= \frac{nh}{r^2 \cdot p^2} C_3 \cdot e^{-\omega t} \cdot \frac{I_n(pr)}{I_n(p\rho)} - \frac{i\omega\varepsilon_m}{p \cdot r} \cdot C_1 \cdot e^{-\omega t} \cdot \frac{I'_n(pr)}{I_n(p\rho)} \\
H_\varphi^{(h)}(r,t) &= \frac{nh}{r^2 \cdot q^2} C_4 \cdot e^{-\omega t} \cdot \frac{K_n(pr)}{K_n(p\rho)} - \frac{i\omega\varepsilon_h}{q \cdot r} \cdot C_2 \cdot e^{-\omega t} \cdot \frac{K'_n(qr)}{K_n(q\rho)} \\
E_\varphi^{(m)}(r,t) &= \frac{nh}{r^2 \cdot p^2} C_1 \cdot e^{-\omega t} \cdot \frac{I_n(pr)}{I_n(p\rho)} + \frac{i\omega}{r \cdot p} C_3 \cdot e^{-\omega t} \cdot \frac{I'_n(pr)}{I_n(p\rho)} \\
E_\varphi^{(h)}(r,t) &= \frac{nh}{r^2 \cdot q^2} C_2 \cdot e^{-\omega t} \cdot \frac{K_n(qr)}{K_n(q\rho)} + \frac{i\omega}{r \cdot q} C_4 \cdot e^{-\omega t} \cdot \frac{K'_n(qr)}{K_n(q\rho)}
\end{aligned} \tag{3.6}$$

Условие непрерывности тангенциальных составляющих полей на границе цилиндра $r = \rho$ приводит к системе уравнений:

$$\frac{nh}{p^2} \cdot C_3 - \frac{i\omega\varepsilon_m}{p^2} \cdot D_I \cdot C_1 = \frac{nh}{q^2} \cdot C_4 - \frac{i\omega\varepsilon_h}{q^2} \cdot D_K \cdot C_2$$

$$\frac{nh}{p^2} \cdot C_1 + \frac{i\omega}{p^2} \cdot D_I \cdot C_3 = \frac{nh}{q^2} \cdot C_2 + \frac{i\omega}{q^2} \cdot D_K \cdot C_4 \qquad (3.7)$$

$$C_1 = C_2$$

$$C_3 = C_4$$

где введены обозначения

$$D_I = \frac{p \cdot \rho \cdot I'_n(p\rho)}{I_n(p\rho)} \qquad\qquad D_K = \frac{q \cdot \rho \cdot K'_n(q\rho)}{K_n(q\rho)} \qquad (3.8)$$

Равенство нулю определителя системы (3.7) приводит к характеристическому уравнению (см. Приложение В):

$$\left(\varepsilon_h \cdot p^2 \cdot D_K - \varepsilon_m \cdot q^2 \cdot D_I\right) \cdot \left(p^2 \cdot D_K - q^2 \cdot D_I\right) - n^2 h^2 \cdot \omega^2 \cdot (\varepsilon_m - \varepsilon_h)^2 = 0 \qquad (3.9)$$

Особо следует рассматривать случай $n = 0$, тогда из (3.9) видно, что

$$\varepsilon_h \cdot \frac{K'_0(q\rho)}{q \cdot K_0(q\rho)} - \varepsilon_m \cdot \frac{I'_0(p\rho)}{p \cdot I_0(p\rho)} = 0 \qquad (3.10)$$

4. Итак, для каждой пары значений $n, k$ с помощью характеристического уравнения (3.9) определяется частота поверхностной волны $\omega$, а затем по формулам (3.2) и (3.3) вычисляются величины $p, q$, определяющие зависимость напряженностей поля от расстояния до оси цилиндра. Поскольку частота $\omega$ входит в (3.9) в квадрате, то частота может принимать два значение — положительное и отрицательное, что соответствует волнам, бегущим в сторону положительных и отрицательных $z$.

При выполнении (3.9) система уравнений имеет решение, зависящее от одной произвольной комплексной константы. Вычислив величины $C_1, C_2, C_3, C_4$ несложно найти компоненты полей $\vec{E}_{nk}, \vec{H}_{nk}$ (см. Приложение C):

Из (3.5) находим:

$$E^{(m)}_{z,nk}(r) = e_z \cdot I_n(pr)/I_n(p\rho) \cdot \gamma$$

$$E^{(h)}_{z,nk}(r) = e_z \cdot K_n(qr)/K_n(q\rho) \cdot \gamma \qquad (4.1)$$

$$H^{(m)}_{z,nk}(r) = h_z \cdot I_n(pr)/I_n(p\rho) \cdot \gamma$$

$$H^{(h)}_{z,nk}(r) = h_z \cdot K_n(qr)/K_n(q\rho) \cdot \gamma$$

Где

$$e_z = \left(q^2 \cdot D_I - p^2 \cdot D_K\right)$$

$$h_z = -i \cdot nh \cdot \omega \cdot (\varepsilon_h - \varepsilon_m) \qquad (4.2)$$

Для азимутальной и радиальной компонент вычисления приводят к выражениям:

$$H_{\varphi,nk}^{(m)}(r) = \left( \frac{nh}{r^2 \cdot p^2} \cdot h_z \cdot \frac{I_n(pr)}{I_n(p\rho)} - \frac{i\omega\varepsilon_m}{p \cdot r} \cdot e_z \cdot \frac{I'_n(pr)}{I_n(p\rho)} \right) \cdot \gamma$$

$$H_{\varphi,nk}^{(h)}(r) = \left( \frac{nh}{r^2 \cdot q^2} \cdot h_z \cdot \frac{K_n(qr)}{K_n(q\rho)} - \frac{i\omega\varepsilon_h}{q \cdot r} \cdot e_z \cdot \frac{K'_n(qr)}{K_n(q\rho)} \right) \cdot \gamma \quad (4.3)$$

$$E_{\varphi,nk}^{(m)}(r) = \left( \frac{nh}{r^2 \cdot p^2} \cdot e_z \cdot \frac{I_n(pr)}{I_n(p\rho)} + \frac{i\omega}{r \cdot p} \cdot h_z \cdot \frac{I'_n(pr)}{I_n(p\rho)} \right) \cdot \gamma$$

$$E_{\varphi,nk}^{(h)}(r) = \left( \frac{nh}{r^2 \cdot q^2} \cdot e_z \cdot \frac{K_n(qr)}{K_n(q\rho)} + \frac{i\omega}{r \cdot q} \cdot h_z \cdot \frac{K'_n(qr)}{K_n(q\rho)} \right) \cdot \gamma$$

$$E_{r,nk}^{(m)} = \left( \frac{n \cdot \omega}{r \cdot p^2} \cdot h_z \cdot \frac{I_n(pr)}{I_n(p\rho)} - \frac{ih}{p} \cdot e_z \cdot \frac{I'_n(pr)}{I_n(p\rho)} \right) \cdot \gamma$$

$$E_{r,nk}^{(h)} = \left( \frac{n\omega}{r \cdot q^2} \cdot h_z \cdot \frac{K_n(qr)}{K_n(q\rho)} - \frac{ih}{q} \cdot e_z \cdot \frac{K'_n(qr)}{K_n(q\rho)} \right) \cdot \gamma \quad (4.4)$$

$$H_{r,nk}^{(m)} = \left( -\frac{n \cdot \varepsilon_m \omega}{r \cdot p^2} \cdot e_z \cdot \frac{I_n(pr)}{I_n(p\rho)} - \frac{ih}{p} \cdot h_z \cdot \frac{I'_n(pr)}{I_n(p\rho)} \right) \cdot \gamma$$

$$H_{r,nk}^{(h)} = \left( -\frac{n \cdot \varepsilon_h \omega}{r \cdot q^2} \cdot e_z \cdot \frac{K_n(qr)}{K_n(q\rho)} - \frac{ih}{q} \cdot h_z \cdot \frac{K'_n(qr)}{K_n(q\rho)} \right) \cdot \gamma$$

Определим вектор-потенциалы поля:

| $\vec{H} = rot\vec{A}$ | (4.5) |
|---|---|
| $\vec{E} = -\partial_t \vec{A} - \nabla \varphi$ | (4.6) |

Выберем калибровку так, чтобы $\varphi = 0$. Обеспечим выполнение равенства (4.6), выбрав потенциал для компоненты $n, k$ частоты $\omega$ в виде:

| $\vec{A}_{nk} = \dfrac{1}{i\omega} \vec{E}_{nk}$ | (4.7) |
|---|---|

С помощью (2.1) легко видеть, что и (4.5) также оказывается выполненным.

5. Обратимся теперь к вычислению энергии, переносимой каждой модой. Плотность энергии и поток энергии поля выражаются известными формулами:

$$w = \frac{1}{8\pi} \cdot \left( \varepsilon E^2 + H^2 \right) \qquad \vec{S} = \frac{1}{4\pi} \cdot \left[ \vec{E} \times \vec{H} \right] \quad (5.1)$$

При этом, как легко убедиться из уравнений Максвелла (2.1) имеет место закон сохранения энергии:

$$\frac{d}{dt}\iiint_V w \cdot dV = -\oiint_\Sigma \vec{S}\cdot\vec{n}\cdot d\sigma \tag{5.2}$$

Где $\Sigma$ -- поверхность, ограничивающая объем $V$.

Для вычисления полной энергии, переносимой рассматриваемой волной в расчете на участок цилиндра длины $L$ вычислим интеграл по объему из (5.2).

$$W = \iiint_V w\cdot dV = \frac{1}{4\pi}\cdot\int_0^L dz\int_0^{2\pi}d\varphi\int_0^\infty rdr(\varepsilon E^2 + H^2) \tag{5.3}$$

Где $\vec{E},\vec{H}$ должны быть выражены с помощью разложения (2.3). При интегрировании по координате $dz$ произведений вида $\exp(ihz)\cdot\exp(ih'z)$ ненулевой вклад дадут только члены с $h'=-h$. Аналогично, при интегрировании по $d\varphi$ следует суммировать произведения с $n'=-n$. После чего получаем:

$$W = \frac{L}{4}\cdot\sum_{n=-\infty}^\infty\sum_{k=-\infty}^\infty\int_0^\infty rdr(\varepsilon\vec{E}_{nk}\vec{E}^*_{nk} + \vec{H}_{nk}\vec{H}^*_{nk}) \tag{5.4}$$

Здесь принято во внимание, что поскольку $\vec{E},\vec{H}$ величины вещественные, то

$$\vec{E}_{n,k}(r,t) = \vec{E}^*_{-n,-k}(r,t),\ \vec{H}_{n,k}(r,t) = \vec{H}^*_{-n,-k}(r,t) \tag{5.5}$$

Что легко получить перейдя к комплексно сопряженным выражениям в (2.3). Характеристическое уравнение для каждой пары $n,k$ имеет два корня $+\omega,-\omega$. Эти значения физически соответствуют волнам, бегущим в положительном и отрицательном направлении оси $z$, соответственно. Мы будем считать, что каждой паре значений $n,k$ соответствуют две независимые моды. Таким образом, энергия распадается на сумму энергий, переносимых каждой модой:

$$W = \sum_{n=-\infty}^\infty\sum_{k=-\infty}^\infty W_{nk} \tag{5.5}$$

где

$$W_{nk} = \frac{L}{4}\cdot\left(\int_0^\rho rdr(\varepsilon_m\vec{E}^{(m)}_{nk}\vec{E}^{(m)*}_{nk} + \vec{H}^{(m)}_{nk}\vec{H}^{(m)*}_{nk}) + \int_\rho^\infty rdr(\varepsilon_h\vec{E}^{(h)}_{nk}\vec{E}^{(h)*}_{nk} + \vec{H}^{(h)}_{nk}\vec{H}^{(h)*}_{nk})\right) \tag{5.6}$$

6. Интегралы (5.6) вычислены в Приложениях Е1, Е2.

$$\begin{aligned}W^{(m)}_{nk} = &-\frac{L}{4}\cdot(\varepsilon_m\cdot e_z\cdot e_z^* + h_z\cdot h_z^*)\cdot\frac{\varepsilon_m\cdot\omega^2\cdot\rho^2}{p^2}\cdot\left(1+\frac{n^2-D_I^2-2\cdot D_I}{p^2\cdot\rho^2}\right)\cdot\gamma_{nk}\cdot\gamma^*_{nk} + \\ &+\frac{L}{4}\cdot(\varepsilon_m\cdot e_z\cdot e_z^* + h_z\cdot h_z^*)\frac{1}{p^2}\cdot D_I\cdot\gamma_{nk}\cdot\gamma^*_{nk} + \frac{L}{2}\cdot(h_z\cdot e_z^* - h_z^*\cdot e_z)\cdot\frac{inh\cdot\varepsilon_m\omega}{p^4}\cdot\gamma_{nk}\cdot\gamma^*_{nk}\end{aligned} \tag{6.1}$$

$$\begin{aligned}W^{(h)}_{nk} = &\frac{L}{4}\cdot(\varepsilon_h\cdot e_z\cdot e_z^* + h_z\cdot h_z^*)\cdot\frac{\varepsilon_h\cdot\omega^2\cdot\rho^2}{q^2}\cdot\left(1+\frac{n^2-D_K^2-2D_K}{q^2\rho^2}\right)\cdot\gamma_{nk}\cdot\gamma^*_{nk} \\ &-\frac{L}{4}\cdot(\varepsilon_h\cdot e_z\cdot e_z^* + h_z\cdot h_z^*)\cdot\frac{D_K}{q^2}\cdot\gamma_{nk}\cdot\gamma^*_{nk} + \frac{L}{2}\cdot(e_z^*\cdot h_z - e_z\cdot h_z^*)\cdot\frac{inh\cdot\varepsilon_h\omega}{q^4}\cdot\gamma_{nk}\cdot\gamma^*_{nk}\end{aligned} \tag{6.2}$$

Полная энергия, переносимая волной равная сумме выражений (6.1) и (6.2).

Определим канонические переменные

$$q_{nk} = (\gamma_{nk} \cdot e^{-i\omega t} + \gamma_{nk}^* \cdot e^{i\omega t})\frac{\sqrt{\Theta_{nk}}}{\omega} \qquad p_{nk} = -i(\gamma_{nk} \cdot e^{-i\omega t} - \gamma_{nk}^* \cdot e^{i\omega t})\sqrt{\Theta_{nk}} \qquad (6.3)$$

где величина $\Theta_{nk}$ выбирается так, чтобы

$$W_{nk}^{(m)} + W_{nk}^{(h)} = 2 \cdot \Theta_{nk} \cdot \gamma_{nk} \cdot \gamma_{nk}^* \qquad (6.4)$$

Очевидно, что

$$\dot{q}_{nk} = p_{nk} \qquad \dot{p}_{nk} = -\omega_{nk}^2 q_{nk} \qquad (6.5)$$

Теперь энергия, колебаний на данной моде выражается через $p_{nk}, q_{nk}$:

$$W_{nk} = \frac{1}{2}\left(\omega_{nk}^2 \cdot q_{nk}^2 + p_{nk}^2\right) \qquad (6.6)$$

Функция Гамильтона системы поверхностных волн имеет вид:

$$H = \frac{1}{2}\sum_{n,k}\left(p_{nk}^2 + \omega_{nk}^2 \cdot q_{nk}^2\right) \qquad (6.7)$$

7. При переходе к квантовой теории мы должны рассматривать канонические переменные $p_{nk}, q_{nk}$ как операторы с правилом коммутации:

$$\widehat{p}_{nk}\widehat{q}_{nk} - \widehat{q}_{nk}\widehat{p}_{nk} = -i\hbar \qquad (7.1)$$

Теперь мы можем выразить через операторы $\widehat{q}_k, \widehat{p}_k$ компоненты напряженности и потенциала электромагнитного поля. Учитывая (6.3) находим:

$$\widehat{\vec{E}}_{nk}^{(u)}(t,z) = \frac{\vec{e}_{nk}^{(u)}(r)}{2\sqrt{\Theta_{nk}}} \cdot (i\widehat{p}_{nk} + \omega_{nk}\widehat{q}_{nk})$$

$$\widehat{\vec{H}}_{nk}^{(u)}(t,z) = \frac{\vec{h}_{nk}^{(u)}(r)}{2\sqrt{\Theta_{kn}}} \cdot (i\widehat{p}_{nk} + \omega_{nk}\widehat{q}_{nk}) \qquad (7.2)$$

$$\widehat{\vec{A}}_{nk}^{(u)}(t,r) = \frac{\vec{a}_{nk}^{(u)}(r)}{2\sqrt{\Theta_{nk}}} \cdot (i\widehat{p}_{nk} + \omega_{nk}\widehat{q}_{nk}) \qquad (7.3)$$

где величины $\vec{e}_{nk}^{(u)}(r)$, $\vec{h}_{nk}^{(u)}(r)$ вычисляются из (4.1)-(4.4). Ниже нам понадобится значение потенциала вне стержня. Найдем $\vec{a}_{nk}^{(h)}(r)$ с помощью (4.7):

$$a_{r,nk}^{(h)}(r) = \frac{1}{i\omega} \cdot \left(\frac{n\omega}{r \cdot q^2} \cdot h_z \cdot \frac{K_n(qr)}{K_n(q\rho)} - \frac{ih}{q} \cdot e_z \cdot \frac{K'_n(qr)}{K_n(q\rho)}\right)$$

$$a_{\varphi,nk}^{(h)}(r) = \frac{1}{i\omega} \cdot \left(\frac{nh}{r^2 \cdot q^2} \cdot e_z \cdot \frac{K_n(qr)}{K_n(q\rho)} + \frac{i\omega}{r \cdot q} \cdot h_z \cdot \frac{K'_n(qr)}{K_n(q\rho)}\right) \qquad (7.4)$$

$$a_{z,nk}^{(h)}(r) = \frac{e_z}{i\omega} \cdot K_n(qr)/K_n(q\rho)$$

8. Гамильтониан системы плазмонов имеет вид вполне аналогичный (6.7):

$$\widehat{H} = \frac{1}{2}\sum_k\left(\widehat{p}_k^2 + \omega_k^2 \cdot \widehat{q}_k^2\right) \qquad (8.1)$$

Гамильтониан распадается на сумму гармонических осцилляторов. Собственные значения такого гамильтониана хорошо известны:

$$E_n = \left(n + \frac{1}{2}\right) \cdot \hbar \omega_k \tag{8.2}$$

А для матричных элементов операторов $\hat{q}_k, \hat{p}_k$ переходов между собственными состояниями гамильтониана можно получить соотношения:

$$\langle n_k | \hat{q}_k | n_k - 1 \rangle = \langle n_k - 1 | \hat{q}_k | n_k \rangle = \sqrt{\frac{\hbar}{2\omega_k} \cdot n_k}$$

$$\langle n_k | \hat{p}_k | n_k - 1 \rangle = -\langle n_k - 1 | \hat{p}_k | n_k \rangle = i\sqrt{\frac{\hbar \omega_k}{2} \cdot n_k} \tag{8.3}$$

Вместо канонических операторов $\hat{q}_k, \hat{p}_k$ удобно использовать их линейные комбинации, имеющие только по одному отличному от нуля матричному элементу для переходов между собственными состояниями гамильтониана (6.4):

$$\hat{c}_k = \frac{1}{\sqrt{2\hbar \omega_k}} \cdot (\omega_k \hat{q}_k + i\hat{p}_k) \qquad \hat{c}_k^+ = \frac{1}{\sqrt{2\hbar \omega_k}} \cdot (\omega_k \hat{q}_k - i\hat{p}_k) \tag{8.4}$$

Несложно показать, что

$$\langle n_k - 1 | \hat{c}_k | n_k \rangle = \sqrt{n_k} \qquad \langle n_k | \hat{c}_k^+ | n_k - 1 \rangle = \sqrt{n_k} \tag{8.5}$$

Правило коммутации для $\hat{c}_k, \hat{c}_k^+$ принимает вид:

$$\hat{c}_k \hat{c}_k^+ - \hat{c}_k^+ \hat{c}_k = 1 \tag{8.6}$$

Таким образом мы можем рассматривать операторы $\hat{c}_k, \hat{c}_k^+$ как операторы уничтожения и рождения плазмона.

Оператор вторично квантованного потенциала поля плазмона принимает вид:

$$\hat{\vec{A}}^{(h)} = \sum_{nk} \sqrt{\frac{\hbar \omega_{nk}}{2\Theta_{nk}}} \left( \hat{c} \cdot \vec{a}_{nk}^{(h)}(r) \cdot \exp(ihz + in\varphi - i\omega t) + \hat{c}^+ \cdot \vec{a}_{nk}^{(h)*}(r) \cdot \exp(-ihz - in\varphi + i\omega t) \right) \tag{8.6}$$

9. Вычислим вероятность излучения плазмона в единицу времени излучателем, расположенным вблизи металлического цилиндра. Размеры излучателя будем считать малыми, сравнительно с длиной волны плазмона, и вероятность вычислим в дипольном приближении. Будем следовать методу расчета вероятности спонтанного излучения, изложенному в /8,§45/ для случая излучателя в свободном пространстве.

Взаимодействие электромагнитного поля с зарядом описывается оператором:

$$\hat{V} = e \cdot \iiint dV \cdot \hat{A}_\mu \hat{j}^\mu \tag{9.1}$$

Здесь $e$ — заряд электрона, $\hat{A}_\mu, \hat{j}^\mu$ — вторично квантованные операторы потенциала электромагнитного поля и "плотности тока электрона". Плотность тока выражается через оператор волновой функции и матрицы Дирака:

$$\hat{j}^\mu = \hat{\overline{\psi}} \gamma^\mu \hat{\psi} \tag{9.2}$$

Интегрирование в (9.1) выполняется по всему 3-пространству.

Согласно теории возмущений вероятность перехода в единицу времени системы из состояния, описываемого набором квантовых чисел *i*, в состояние *f* выражается формулой:

$$w_{fi} = \frac{2\pi}{\hbar}\left|V_{fi}\right|^2 \cdot \delta(E_f - E_i - \hbar\omega) \qquad (9.3)$$

Матричный элемент оператора возмущения $V_{fi}$ вычисляется с помощью невозмущенных волновых функций состояний *i* и *f*. Будем считать для простоты, что как излучатель, так и поле имеют дискретный спектр состояний.

В дипольном приближении считают величины $A_\mu(x,y,z)$ медленно меняющимися в пределах характерных размеров излучателя и выносят их из-под знака интеграла (9.1). В результате матричный элемент оператора (9.1) приобретает вид:

$$\left\langle 1_{nk} f \left| \hat{V} \right| 0_{nk} i \right\rangle = -e \cdot \left\langle 1_{nk} \left| \hat{\vec{A}}_\kappa(\vec{r}_{em}) \right| 0_{nk} \right\rangle \cdot \iiint dV \cdot \psi_f^* \gamma^0 \vec{\gamma} \psi_i \qquad (9.4)$$

здесь $\left\langle 1_{nk} \left| \hat{\vec{A}} \right| 0_{nk} \right\rangle$ — матричный элемент, соответствующий спонтанному излучению:

$$\left\langle 1_{nk} \left| \hat{\vec{A}}(\vec{r}_{em}) \right| 0_{nk} \right\rangle = \sqrt{\frac{\hbar\omega_{nk}}{2\Theta_{nk}}} \cdot \vec{a}_{nk}^{(h)*}(\vec{r}_{em}) \cdot \exp(-ihz - in\varphi + i\omega t) \qquad (9.5)$$

а через $\vec{r}_{em}$ обозначены координаты точки, где находится излучатель. Интеграл по объему в (9.4) в нерелятивистском приближении есть матричный элемент скорости электрона, вычисляемый по нерелятивистским собственным функциям. Матричный элемент скорости связан с матричным элементом координаты и частотой перехода: $\vec{v}_{fi} = -i\omega\vec{r}_{fi}$. Вводя дипольный момент системы $\vec{d} = e\vec{r}$, находим для квадрата модуля матричного элемента:

$$\left|\left\langle 1_{nk} f \left| \hat{V} \right| 0_{nk} i \right\rangle\right|^2 = \left|\vec{d}_{fi} \cdot \vec{a}_{nk}^{(h)*}(\vec{r}_{em})\right|^2 \cdot \frac{\hbar\omega_{nk}^3}{2\Theta_{nk}} \qquad (9.6)$$

10. Будем считать, что излучатели вблизи цилиндра ориентированы хаотично. Усредним выражение (9.6) по направлению дипольного момента излучателя.

$$\overline{\left|\vec{d}_{fi} \cdot \vec{a}_{nk}^{(h)*}(\vec{r}_{em})\right|^2} = \overline{\left(d_x a_x^* + d_y a_y^* + d_z a_z^*\right) \cdot \left(d_x^* a_x + d_y^* a_y + d_z^* a_z\right)} \qquad (10.1)$$

Учитывая, что

$$\overline{d_x d_y^*} = \overline{d_x d_z^*} = \overline{d_y d_z^*} = 0 \qquad \overline{d_x d_x^*} = \overline{d_y d_y^*} = \overline{d_z d_z^*} = \frac{1}{3}\cdot\left|\vec{d}_{fi}\right|^2 \qquad (10.2)$$

Находим для среднего значения:

$$\overline{\left|\vec{d}_{fi} \cdot \vec{a}_{nk}^{(h)*}(\vec{r}_{em})\right|^2} = \frac{1}{3}\cdot\left|d_{fi}\right|^2 \cdot \left|\vec{a}_{nk}^{(h)*}(\vec{r}_{em})\right|^2 \qquad (10.3)$$

Итак, для вероятности перехода получаем:

$$\overline{w_{fi,nk}} = \frac{\pi \cdot \omega^3}{3 \cdot \Theta_{nk}}\left|d_{fi}\right|^2 \cdot \left|\vec{a}_{nk}^{(h)*}(\vec{r}_{em})\right|^2 \cdot \delta(E_f - E_i - \hbar\omega) \qquad (10.4)$$

11. Для вычисления полной вероятности перехода с излучением плазмона просуммируем величину (10.3) по всем модам, определяемым набором $n, k$. Сумму ряда по дискретному набору состояний заменим интегралом, учитывая при этом, что на каждый интервал значений $\Delta h$ приходится $\Delta N_n = \dfrac{\Delta h}{(2\pi/L)}$ состояний поля, где $L$ — длина выбранного отрезка металлического цилиндра.

Таким образом мы приходим к выражению для вероятности спонтанного перехода:

$$\overline{w_{fi,n}} = \int_{-\infty}^{\infty} \frac{L \cdot dh}{2\pi} \cdot \frac{\pi}{3} \cdot \frac{\omega^3}{\Theta_{nk}} \cdot \left|\vec{d}_{fi}\right|^2 \cdot \left|\vec{a}_{nk}^{(h)*}(\vec{r}_{em})\right|^2 \cdot \delta(E_f - E_i - \hbar\omega) \tag{11.2}$$

Переходя в (11.2) к интегрированию по $d\omega$ и вычисляя интеграл δ-функции, найдем

$$\overline{w_{f,n}} = \frac{L}{6\hbar} \cdot \left|\vec{d}_{fi}\right|^2 \cdot \left(\frac{\partial h}{\partial \omega}\right) \cdot \frac{\omega^3}{\Theta_{nk}} \cdot \left|\vec{a}_{nk}^{(h)*}(\vec{r}_{em})\right|^2 \tag{11.3}$$

Сравним полученное выражение с известной формулой вероятности дипольного излучения в свободном пространстве /8,§45/:

$$w_{ph} = \frac{4\omega^3}{3\hbar} \cdot \left|\vec{d}_{fi}\right|^2 \tag{11.4}$$

Введем фактор $F$, характеризующий возрастание вероятности спонтанного излучения плазмона с азимутальным числом $n$ вблизи тонкого металлического цилиндра, относительно вероятности излучения фотона:

$$F_n = \frac{\overline{w_{fi,n}}}{w_{ph}} = \frac{1}{8} \cdot \left(\frac{\partial h}{\partial \omega}\right) \cdot \frac{L}{\Theta_{nk}} \cdot \left|\vec{a}_{nk}^{(h)*}(\vec{r}_{em})\right|^2 \tag{11.5}$$

Ниже приведены дисперсионные кривые $\omega(k)$ и значения величины $F_n$ при различных значениях параметров.

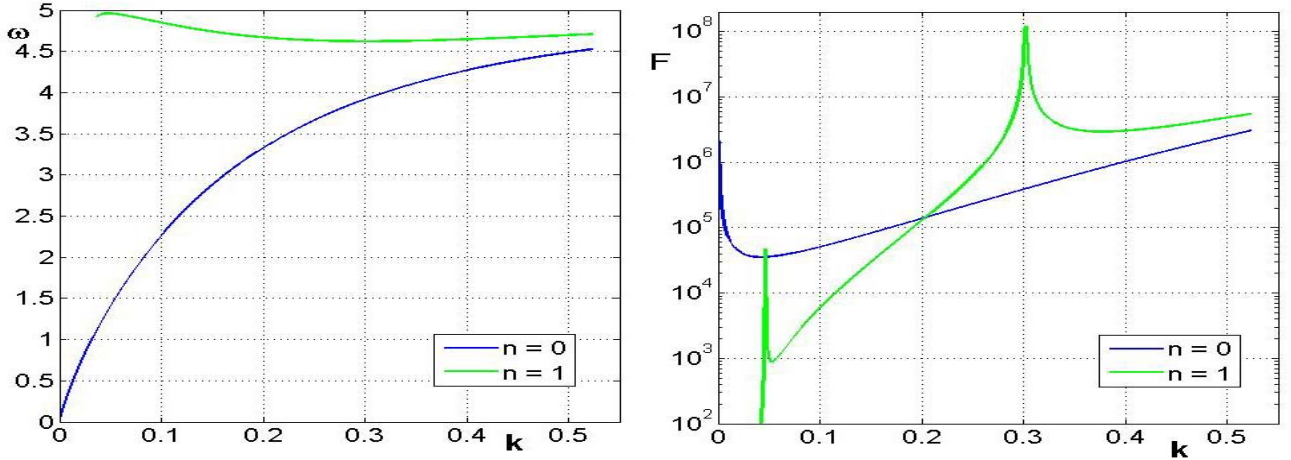

(a)          (b)

Рис. 2. Зависимость от волнового вектора частоты резонансного плазмона (a) и фактора $F$ (b) для цилиндрических гармоник с $n = 0,1$ Волновой вектор в единицах $[nm^{-1}]$, частота в $[eV]$. При расчете приняты значения: $\omega_{pl} = 9.1 eV$; $\varepsilon_h = 2$; $\rho = 5 nm$.

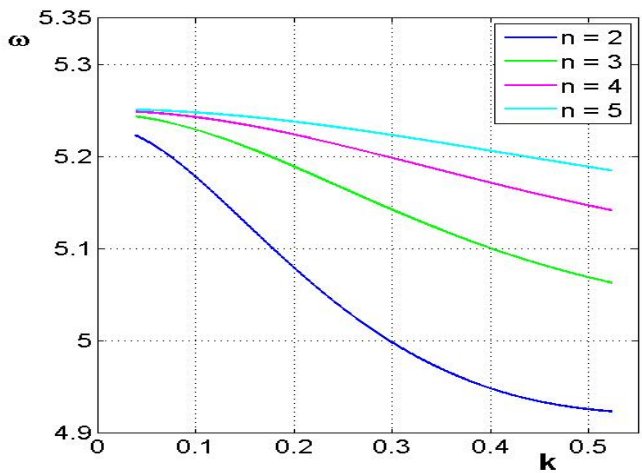 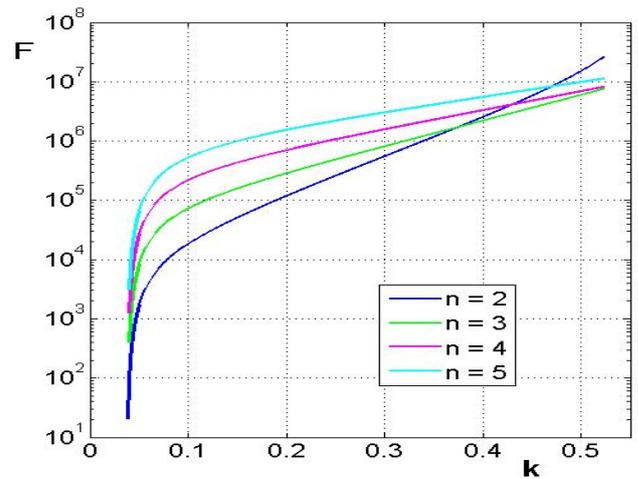

(a) (b)

Рис. 3. Зависимость от волнового вектора частоты резонансного плазмона (a) и фактора $F$ (b) для цилиндрических гармоник с $n = 2,3,4,5$. Волновой вектор в единицах $[nm^{-1}]$, частота в $[eV]$. При расчете приняты значения: $\omega_{pl} = 9.1 eV$; $\varepsilon_h = 2$; $\rho = 5 nm$.

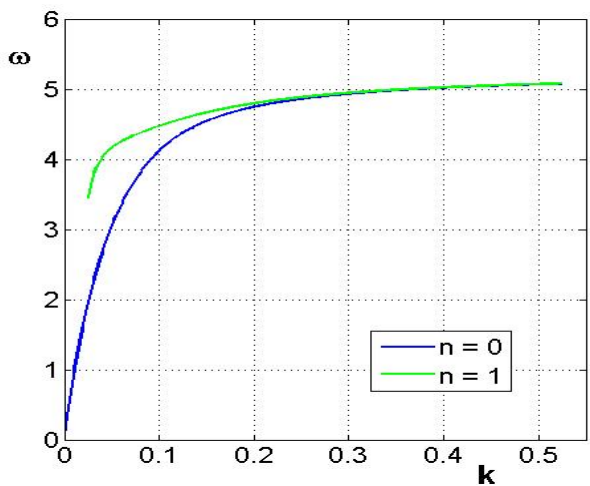 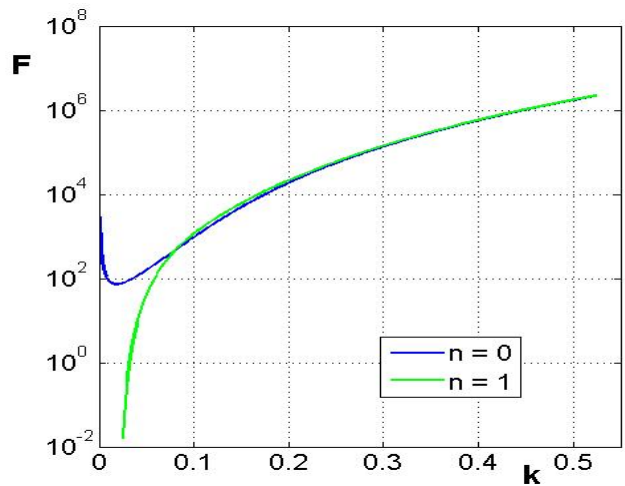

(a) (b)

Рис. 4. Зависимость от волнового вектора частоты резонансного плазмона (a) и фактора $F$ (b) для цилиндрических гармоник с $n = 0,1$. Волновой вектор в единицах $[nm^{-1}]$, частота в $[eV]$. При расчете приняты значения: $\omega_{pl} = 9.1 eV$; $\varepsilon_h = 2$; $\rho = 20 nm$.

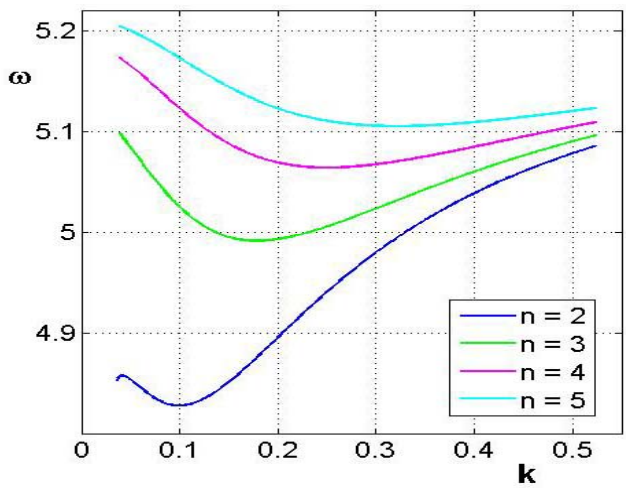 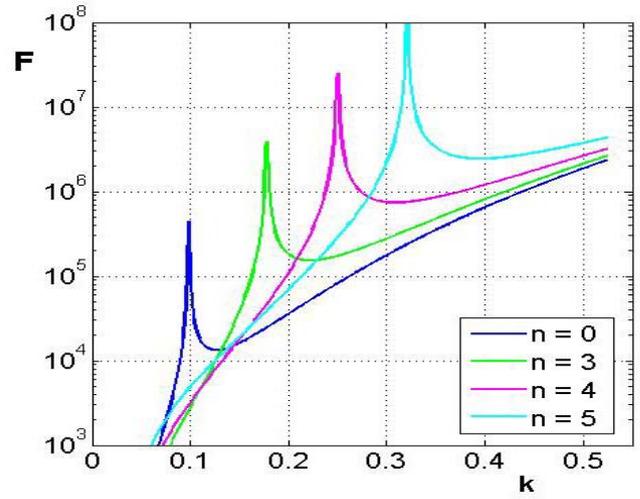

(a)                                          (b)

Рис. 5. Зависимость от волнового вектора частоты резонансного плазмона (a) и фактора $F$ (b) для цилиндрических гармоник с $n = 2, 3, 4, 5$. Волновой вектор в единицах $[nm^{-1}]$, частота в $[eV]$. При расчете приняты значения: $\omega_{pl} = 9.1 eV$; $\varepsilon_h = 2$; $\rho = 20 nm$.

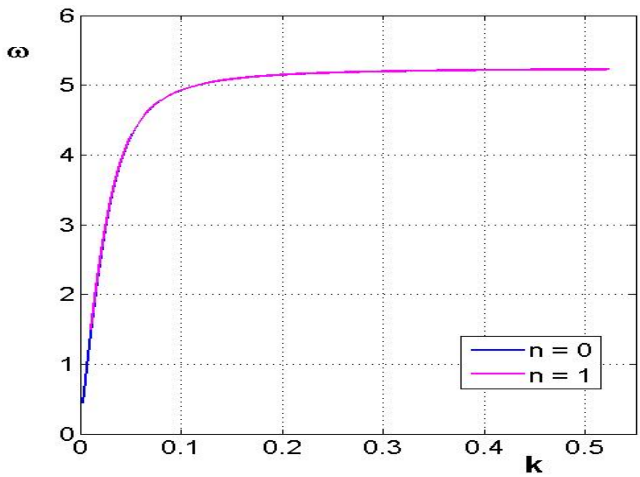 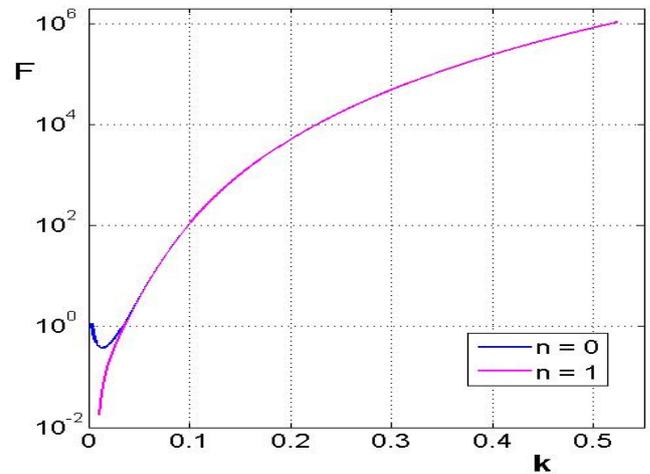

(a)                                          (b)

Рис. 6. Зависимость от волнового вектора частоты резонансного плазмона (a) и фактора $F$ (b) для цилиндрических гармоник с $n = 2, 3, 4, 5$. Волновой вектор в единицах $[nm^{-1}]$, частота в $[eV]$. При расчете приняты значения: $\omega_{pl} = 9.1 eV$; $\varepsilon_h = 2$; $\rho = 200 nm$.

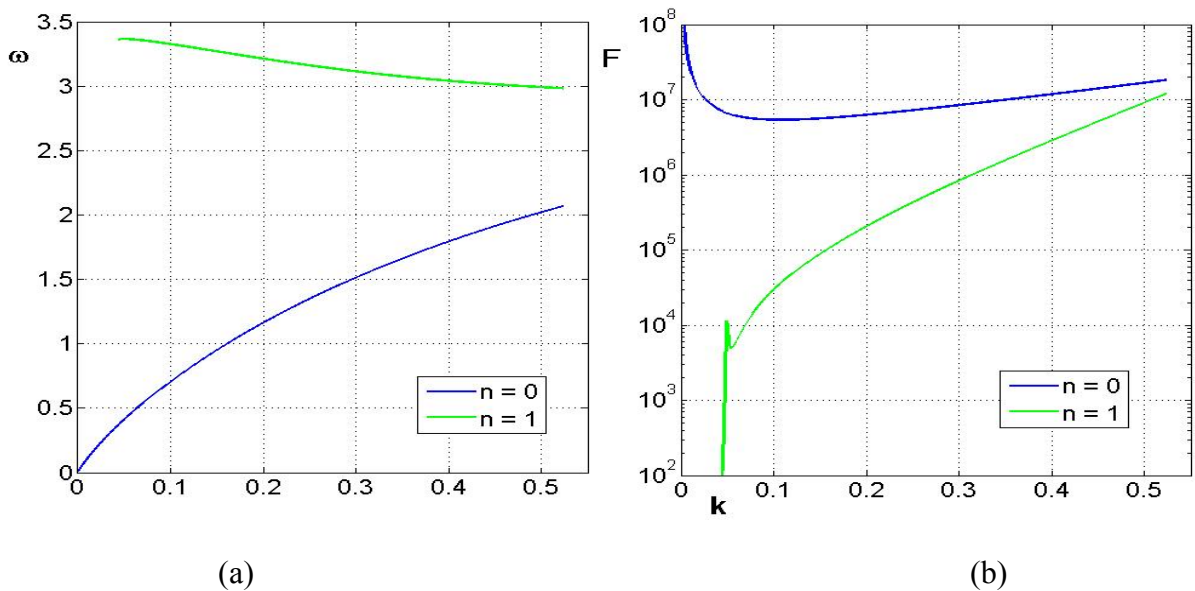

(a) (b)

Рис. 7. Зависимость от волнового вектора частоты резонансного плазмона (a) и фактора $F$ (b) для цилиндрических гармоник с $n = 0,1$. Волновой вектор в единицах $[nm^{-1}]$, частота в $[eV]$. При расчете приняты значения: $\omega_{pl} = 9.1 eV$; $\varepsilon_h = 6$; $\rho = 2 nm$.

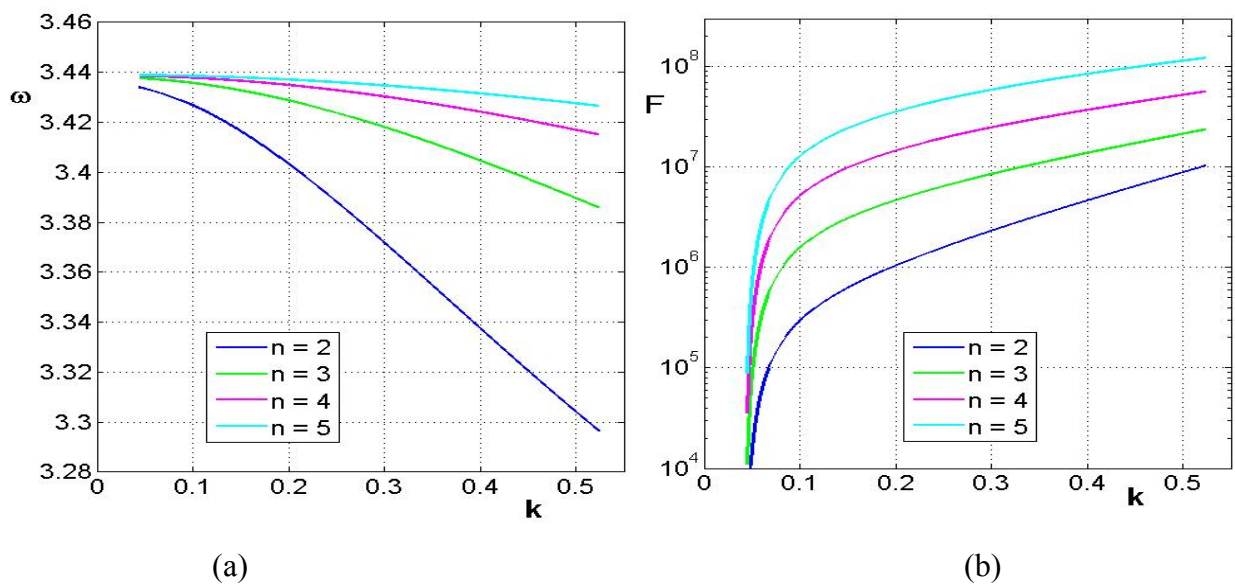

(a) (b)

Рис. 8. Зависимость от волнового вектора частоты резонансного плазмона (a) и фактора $F$ (b) для цилиндрических гармоник с $n = 2,3,4,5$. Волновой вектор в единицах $[nm^{-1}]$, частота в $[eV]$. При расчете приняты значения: $\omega_{pl} = 9.1 eV$; $\varepsilon_h = 6$; $\rho = 2 nm$.

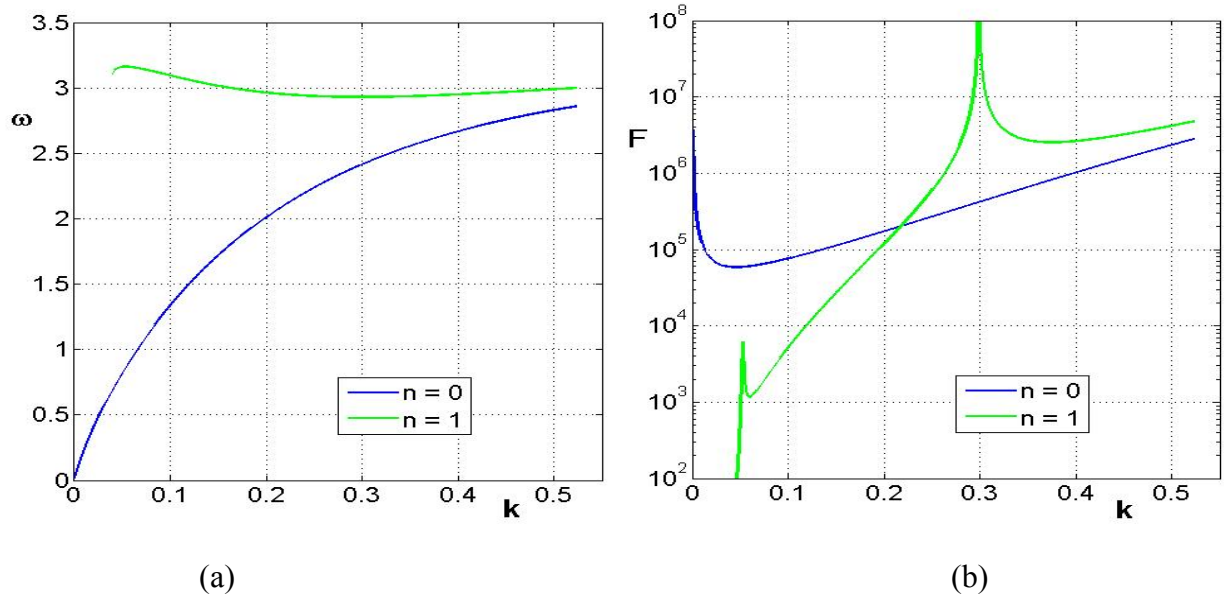

Рис. 9. Зависимость от волнового вектора частоты резонансного плазмона (a) и фактора $F$ (b) для цилиндрических гармоник с $n = 0,1$. Волновой вектор в единицах $[nm^{-1}]$, частота в $[eV]$. При расчете приняты значения: $\omega_{pl} = 9.1 eV$; $\varepsilon_h = 6$; $\rho = 5 nm$.

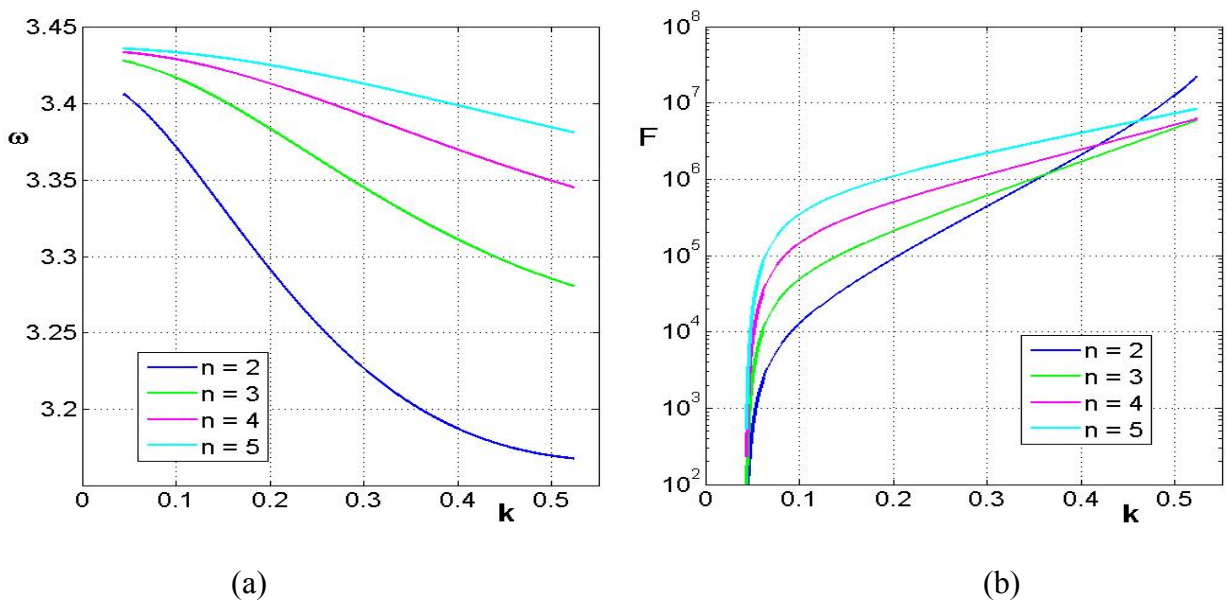

Рис. 10. Зависимость от волнового вектора частоты резонансного плазмона (a) и фактора $F$ (b) для цилиндрических гармоник с $n = 2,3,4,5$. Волновой вектор в единицах $[nm^{-1}]$, частота в $[eV]$. При расчете приняты значения: $\omega_{pl} = 9.1 eV$; $\varepsilon_h = 6$; $\rho = 5 nm$.

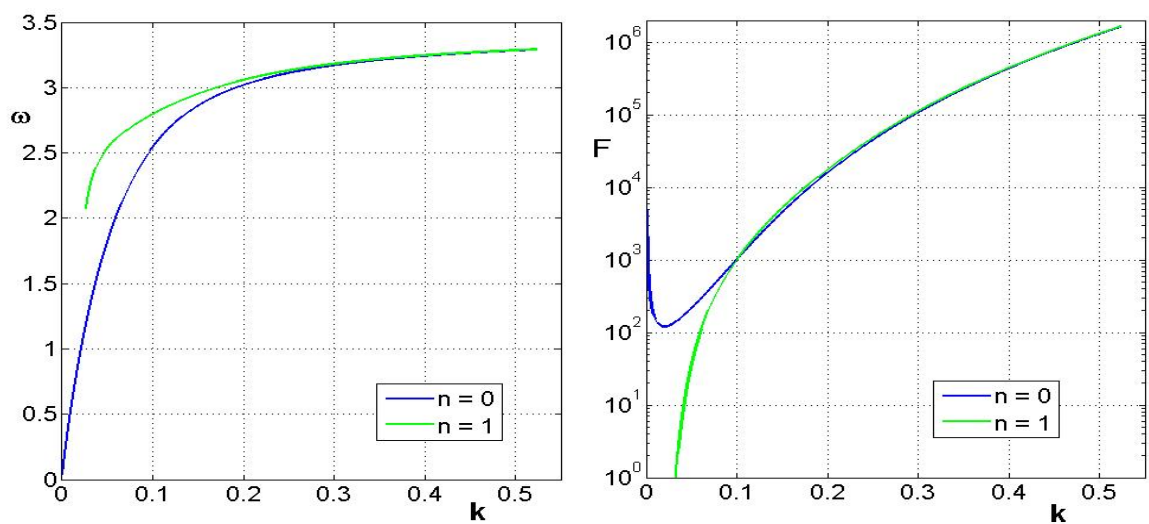

Рис. 11. Зависимость от волнового вектора частоты резонансного плазмона (a) и фактора $F$ (b) для цилиндрических гармоник с $n = 0,1$. Волновой вектор в единицах $[nm^{-1}]$, частота в $[eV]$. При расчете приняты значения: $\omega_{pl} = 9.1 eV$; $\varepsilon_h = 6$; $\rho = 20 nm$.

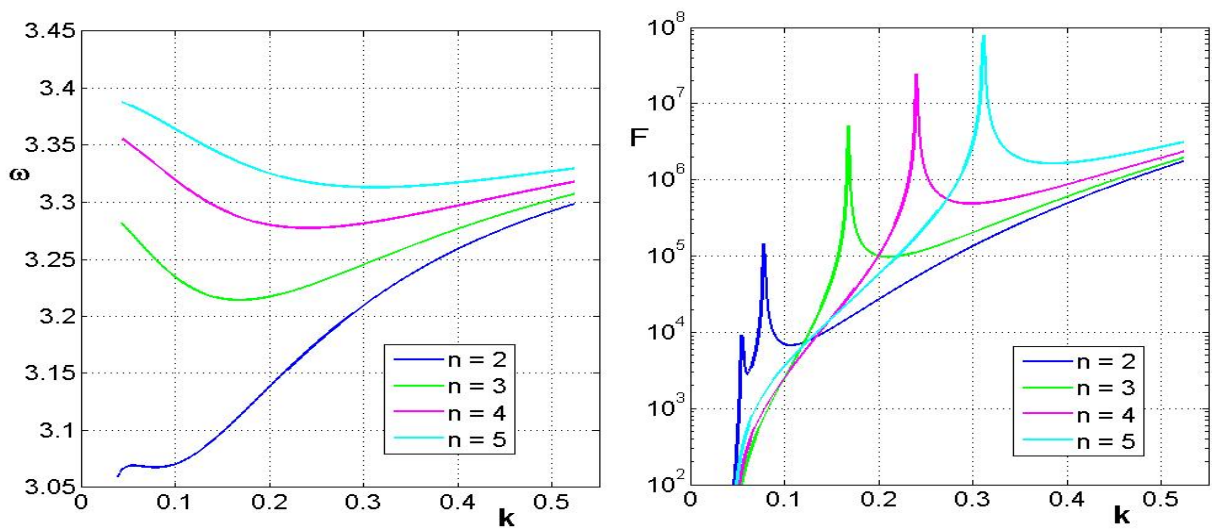

Рис. 12. Зависимость от волнового вектора частоты резонансного плазмона (a) и фактора $F$ (b) для цилиндрических гармоник с $n = 2,3,4,5$. Волновой вектор в единицах $[nm^{-1}]$, частота в $[eV]$. При расчете приняты значения: $\omega_{pl} = 9.1 eV$; $\varepsilon_h = 6$; $\rho = 20 nm$.

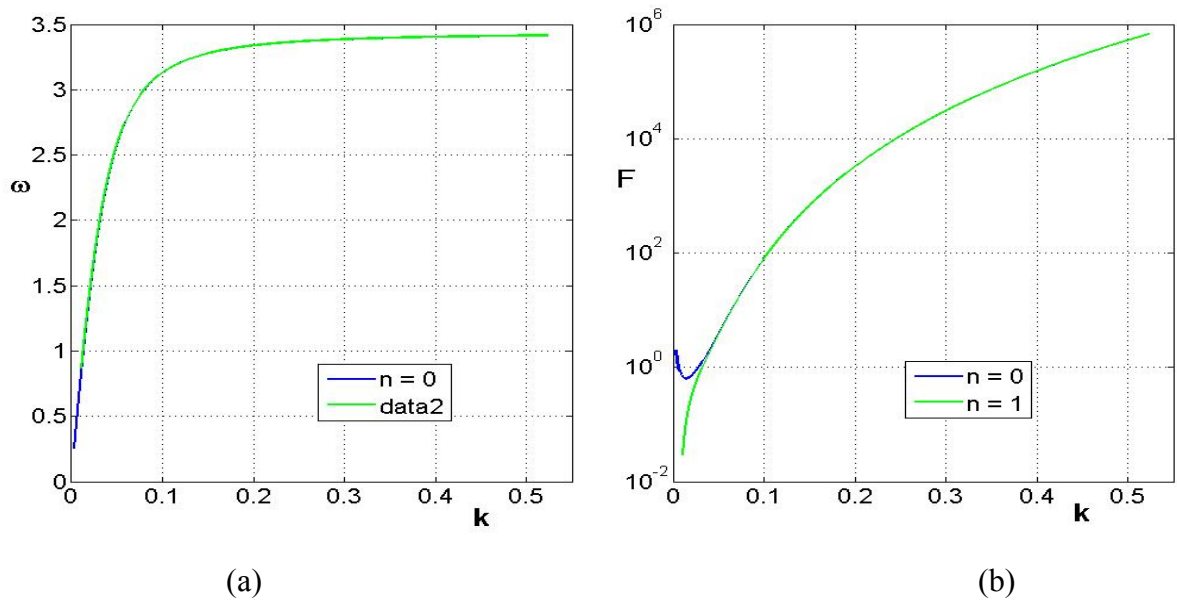

(a)                      (b)

Рис. 13. Зависимость от волнового вектора частоты резонансного плазмона (a) и фактора $F$ (b) для цилиндрических гармоник с $n = 0,1$. Волновой вектор в единицах $[nm^{-1}]$, частота в $[eV]$. При расчете приняты значения: $\omega_{pl} = 9.1 eV$; $\varepsilon_h = 6$; $\rho = 200 nm$.

**Приложение А.**

Из (2.5) и (2.9) находим:

$$\frac{ih}{r}\left(\frac{in}{r}H_z - ihr \cdot H_\varphi\right) \cdot \frac{1}{-i\omega\varepsilon} - \frac{1}{r}\partial_r E_z = i\omega \cdot H_\varphi$$

$$\left(\frac{in}{r}H_z - ihr \cdot H_\varphi\right) \cdot \frac{h}{\omega\varepsilon} + \partial_r E_z = -i\omega r \cdot H_\varphi$$

$$\frac{inh}{r}H_z - ih^2 r \cdot H_\varphi + \omega\varepsilon \cdot \partial_r E_z = -i\omega^2 \varepsilon r \cdot H_\varphi$$

$$\frac{inh}{r}H_z + \omega\varepsilon \cdot \partial_r E_z = i(h^2 - \omega^2\varepsilon) \cdot r \cdot H_\varphi$$

$$H_\varphi = \frac{nh}{r^2 \cdot (h^2 - \omega^2\varepsilon)}H_z - \frac{i\omega\varepsilon}{(h^2 - \omega^2\varepsilon) \cdot r} \cdot \partial_r E_z$$

$$H_\varphi^{(m)} = \frac{nh}{r^2 \cdot p^2}H_z^{(m)} - \frac{i\omega\varepsilon_m}{p^2 \cdot r} \cdot \partial_r E_z^{(m)}$$

$$H_\varphi^{(m)} = \frac{nh}{r^2 \cdot p^2}C_3 \cdot I_n(pr) - \frac{i\omega\varepsilon_m}{p \cdot r}C_1 \cdot I'_n(pr)$$

$$H_\varphi^{(h)} = \frac{nh}{r^2 \cdot q^2}C_4 \cdot K_n(qr) - \frac{i\omega\varepsilon_h}{q \cdot r}C_2 \cdot K'_n(qr)$$

Из (2.6) и (2.8):

$$\frac{ih}{r\omega}\left(\frac{n}{r}E_z - hr \cdot E_\varphi\right) - \frac{1}{r}\partial_r H_z = -i\omega\varepsilon \cdot E_\varphi$$

$$h\left(\frac{n}{r}E_z - hr \cdot E_\varphi\right) + i\omega\partial_r H_z = -\omega^2\varepsilon \cdot rE_\varphi$$

$$\frac{nh}{r}E_z + i\omega\partial_r H_z = \left(h^2 - \omega^2\varepsilon\right) \cdot rE_\varphi$$

$$E_\varphi = \frac{nh}{r^2 \cdot \left(h^2 - \omega^2\varepsilon\right)}E_z + \frac{i\omega}{r \cdot \left(h^2 - \omega^2\varepsilon\right)}\partial_r H_z$$

$$E_\varphi^{(m)} = \frac{nh}{r^2 \cdot p^2}E_z^{(m)} + \frac{i\omega}{r \cdot p^2}\partial_r H_z^{(m)}$$

$$E_\varphi^{(m)} = \frac{nh}{r^2 \cdot p^2}C_1 \cdot I_n(pr) + \frac{i\omega}{r \cdot p}C_3 \cdot I'_n(pr)$$

$$E_\varphi^{(h)} = \frac{nh}{r^2 \cdot q^2} C_2 \cdot K_n(qr) + \frac{i\omega}{r \cdot q} C_4 \cdot K'_n(qr)$$

Из (2.5) и (2.8) находим:

$$\frac{in}{r} H_z - ihr \cdot H_\varphi = -i\omega \cdot E_r$$

$$\frac{in}{r} E_z - ihr \cdot E_\varphi = i\omega \cdot H_r$$

$$E_r = \frac{1}{\omega\varepsilon} \cdot \left(-\frac{n}{r} H_z + hr \cdot H_\varphi\right)$$

$$H_r = \frac{n}{\omega r} E_z - \frac{hr}{\omega} \cdot E_\varphi$$

Подставляя (2.9) и (2.6):

$$\omega\varepsilon \cdot E_r = -\frac{n}{r} H_z + hr \cdot H_\varphi$$

$$\omega H_r = \frac{n}{r} E_z - hr \cdot E_\varphi$$

$$\omega\varepsilon \cdot E_r = -\frac{n}{r} H_z + \frac{hr}{i\omega} \cdot \left(\frac{ih}{r} E_r - \frac{1}{r} \partial_r E_z\right)$$

$$\omega H_r = \frac{n}{r} E_z - \frac{hr}{-i\omega\varepsilon} \cdot \left(\frac{ih}{r} H_r - \frac{1}{r} \partial_r H_z\right)$$

$$i\omega^2 \varepsilon \cdot E_r = -\frac{i\omega n}{r} H_z + ih^2 E_r - h \cdot \partial_r E_z$$

$$-i\varepsilon\omega^2 H_r = \frac{-i\omega\varepsilon n}{r} E_z - \left(ih^2 H_r - h\partial_r H_z\right)$$

$$i(h^2 - \omega^2\varepsilon) \cdot E_r = \frac{i\omega n}{r} H_z + h \cdot \partial_r E_z$$

$$i(h^2 - \varepsilon\omega^2) H_r = \frac{-i\omega\varepsilon n}{r} E_z + h\partial_r H_z$$

## **Приложение В.**

Преобразуем систему (3.7)

$$\left(\frac{nh}{p^2} - \frac{nh}{q^2}\right) \cdot C_3 + \left(\frac{i\omega\varepsilon_h}{q^2} \cdot D_K - \frac{i\omega\varepsilon_m}{p^2} \cdot D_I\right) \cdot C_1 = 0$$

$$\left(\frac{i\omega}{p^2}\cdot D_I - \frac{i\omega}{q^2}\cdot D_K\right)\cdot C_3 + \left(\frac{nh}{p^2} - \frac{nh}{q^2}\right)\cdot C_1 = 0$$

$$nh\cdot(q^2 - p^2)\cdot C_3 + i\omega\cdot(\varepsilon_h\cdot p^2\cdot D_K - \varepsilon_m\cdot q^2\cdot D_I)\cdot C_1 = 0$$

$$i\omega\cdot(q^2\cdot D_I - p^2\cdot D_K)\cdot C_3 + nh\cdot(q^2 - p^2)\cdot C_1 = 0$$

$$nh\cdot\omega\cdot(\varepsilon_m - \varepsilon_h)\cdot C_3 + i\cdot(\varepsilon_h\cdot p^2\cdot D_K - \varepsilon_m\cdot q^2\cdot D_I)\cdot C_1 = 0$$

$$i\cdot(q^2\cdot D_I - p^2\cdot D_K)\cdot C_3 + nh\cdot\omega\cdot(\varepsilon_m - \varepsilon_h)\cdot C_1 = 0$$

$$(\varepsilon_h\cdot p^2\cdot D_K - \varepsilon_m\cdot q^2\cdot D_I)\cdot(p^2\cdot D_K - q^2\cdot D_I) - n^2 h^2\cdot\omega^2\cdot(\varepsilon_m - \varepsilon_h)^2 = 0$$

## **Приложение С.**

Преобразуем второе из уравнений (3.7), с учетом третьего и четвертого

$$\frac{nh}{p^2}\cdot C_1 + \frac{i\omega}{p^2}\cdot D_I\cdot C_3 = \frac{nh}{q^2}\cdot C_1 + \frac{i\omega}{q^2}\cdot D_K\cdot C_3$$

$$nh\cdot(q^2 - p^2)\cdot C_1 + i\omega\cdot(q^2\cdot D_I - p^2\cdot D_K)\cdot C_3 = 0$$

$$nh\cdot\omega\cdot(\varepsilon_h - \varepsilon_m)\cdot C_1 = i\cdot(q^2\cdot D_I - p^2\cdot D_K)\cdot C_3$$

| | |
|---|---|
| $nh\cdot\omega\cdot(\varepsilon_h - \varepsilon_m)\cdot C_1 = i\cdot(q^2\cdot D_I - p^2\cdot D_K)\cdot C_3$ | (C.1) |

Решение (C.1) выберем в виде

| | |
|---|---|
| $C_3 = C_4 = -i\cdot nh\cdot\omega\cdot(\varepsilon_h - \varepsilon_m)\cdot\gamma$<br>$C_1 = C_2 = (q^2\cdot D_I - p^2\cdot D_K)\cdot\gamma$ | (C.2) |

Из (3.5) находим:

| | |
|---|---|
| $E_{z,nk\omega}^{(m)}(r) = e_z\cdot I_n(pr)/I_n(p\rho)\cdot\gamma$<br>$E_{z,nk\omega}^{(h)}(r) = e_z\cdot K_n(qr)/K_n(q\rho)\cdot\gamma$<br>$H_{z,nk\omega}^{(m)}(r) = h_z\cdot I_n(pr)/I_n(p\rho)\cdot\gamma$<br>$H_{z,nk\omega}^{(h)}(r) = h_z\cdot K_n(qr)/K_n(q\rho)\cdot\gamma$ | (C.3) |

Где

| | |
|---|---|
| $e_z = (q^2\cdot D_I - p^2\cdot D_K)$ | (C.4) |

$$h_z = -i \cdot nh \cdot \omega \cdot (\varepsilon_h - \varepsilon_m)$$

Из (3.6) легко находятся компоненты $E_\varphi, H_\varphi$:

$$H_{\varphi,nk\omega}^{(m)}(r) = \left( \frac{nh}{r^2 \cdot p^2} \cdot h_z \cdot \frac{I_n(pr)}{I_n(p\rho)} - \frac{i\omega\varepsilon_m}{p \cdot r} \cdot e_z \cdot \frac{I'_n(pr)}{I_n(p\rho)} \right) \cdot \gamma$$

$$H_{\varphi,nk\omega}^{(h)}(r) = \left( \frac{nh}{r^2 \cdot q^2} \cdot h_z \cdot \frac{K_n(qr)}{K_n(q\rho)} - \frac{i\omega\varepsilon_h}{q \cdot r} \cdot e_z \cdot \frac{K'_n(qr)}{K_n(q\rho)} \right) \cdot \gamma \qquad (С.5)$$

$$E_{\varphi,nk\omega}^{(m)}(r) = \left( \frac{nh}{r^2 \cdot p^2} \cdot e_z \cdot \frac{I_n(pr)}{I_n(p\rho)} + \frac{i\omega}{r \cdot p} \cdot h_z \cdot \frac{I'_n(pr)}{I_n(p\rho)} \right) \cdot \gamma$$

$$E_{\varphi,nk\omega}^{(h)}(r) = \left( \frac{nh}{r^2 \cdot q^2} \cdot e_z \cdot \frac{K_n(qr)}{K_n(q\rho)} + \frac{i\omega}{r \cdot q} \cdot h_z \cdot \frac{K'_n(qr)}{K_n(q\rho)} \right) \cdot \gamma$$

Из (2.5) и (2.8) находим:

$$\frac{in}{r} H_z - ihr \cdot H_\varphi = -i\omega\varepsilon \cdot E_r$$

$$\frac{in}{r} E_z - ihr \cdot E_\varphi = i\omega \cdot H_r$$

$$E_r = -\frac{n}{r \cdot \varepsilon\omega} H_z + \frac{hr}{\varepsilon\omega} \cdot H_\varphi$$

$$H_r = \frac{n}{r\omega} E_z - \frac{hr}{\omega} \cdot E_\varphi \qquad (С.6)$$

$$E_r^{(m)} = \left( -\frac{n}{r \cdot \varepsilon_m \omega} \cdot h_z \cdot \frac{I_n(pr)}{I_n(p\rho)} + \frac{hr}{\varepsilon_m \omega} \cdot \left( \frac{nh}{r^2 \cdot p^2} \cdot h_z \cdot \frac{I_n(pr)}{I_n(p\rho)} - \frac{i\omega\varepsilon_m}{p \cdot r} \cdot e_z \cdot \frac{I'_n(pr)}{I_n(p\rho)} \right) \right) \cdot \gamma$$

$$E_r^{(m)} = \left( -\frac{n}{r \cdot \varepsilon_m \omega} \cdot h_z \cdot \frac{I_n(pr)}{I_n(p\rho)} + \frac{hr}{\varepsilon_m \omega} \cdot \frac{nh}{r^2 \cdot p^2} \cdot h_z \cdot \frac{I_n(pr)}{I_n(p\rho)} - \frac{ih}{p} \cdot e_z \cdot \frac{I'_n(pr)}{I_n(p\rho)} \right) \cdot \gamma$$

$$E_r^{(m)} = \left( \left( -1 + \frac{h^2}{p^2} \right) \cdot \frac{n}{r \cdot \varepsilon_m \omega} \cdot h_z \cdot \frac{I_n(pr)}{I_n(p\rho)} - \frac{ih}{p} \cdot e_z \cdot \frac{I'_n(pr)}{I_n(p\rho)} \right) \cdot \gamma$$

$$E_r^{(m)} = \left( \frac{n \cdot \omega}{r \cdot p^2} \cdot h_z \cdot \frac{I_n(pr)}{I_n(p\rho)} - \frac{ih}{p} \cdot e_z \cdot \frac{I'_n(pr)}{I_n(p\rho)} \right) \cdot \gamma$$

$$E_r^{(h)} = \left( -\frac{n}{r \cdot \varepsilon_h \omega} h_z \cdot \frac{K_n(qr)}{K_n(q\rho)} + \frac{hr}{\varepsilon_h \omega} \cdot \left( \frac{nh}{r^2 \cdot q^2} \cdot h_z \cdot \frac{K_n(qr)}{K_n(q\rho)} - \frac{i\omega\varepsilon_h}{q \cdot r} \cdot e_z \cdot \frac{K'_n(qr)}{K_n(q\rho)} \right) \right) \cdot \gamma$$

$$E_r^{(h)} = \left( -\frac{n}{r \cdot \varepsilon_h \omega} h_z \cdot \frac{K_n(qr)}{K_n(q\rho)} + \frac{hr}{\varepsilon_h \omega} \cdot \frac{nh}{r^2 \cdot q^2} \cdot h_z \cdot \frac{K_n(qr)}{K_n(q\rho)} - \frac{ih}{q} \cdot e_z \cdot \frac{K'_n(qr)}{K_n(q\rho)} \right) \cdot \gamma$$

$$E_r^{(h)} = \left( \left( -1 + \frac{h^2}{q^2} \right) \cdot \frac{n}{r \cdot \varepsilon_h \omega} \cdot h_z \cdot \frac{K_n(qr)}{K_n(q\rho)} - \frac{ih}{q} \cdot e_z \cdot \frac{K'_n(qr)}{K_n(q\rho)} \right) \cdot \gamma$$

$$E_r^{(h)} = \left( \frac{n\omega}{r \cdot q^2} \cdot h_z \cdot \frac{K_n(qr)}{K_n(q\rho)} - \frac{ih}{q} \cdot e_z \cdot \frac{K'_n(qr)}{K_n(q\rho)} \right) \cdot \gamma$$

$$H_r^{(m)} = \left( \frac{n}{r\omega} e_z \cdot \frac{I_n(pr)}{I_n(p\rho)} - \frac{hr}{\omega} \cdot \left( \frac{nh}{r^2 \cdot p^2} \cdot e_z \cdot \frac{I_n(pr)}{I_n(p\rho)} + \frac{i\omega}{r \cdot p} \cdot h_z \cdot \frac{I'_n(pr)}{I_n(p\rho)} \right) \right) \cdot \gamma$$

$$H_r^{(m)} = \left( \frac{n}{r\omega} e_z \cdot \frac{I_n(pr)}{I_n(p\rho)} - \frac{hr}{\omega} \cdot \frac{nh}{r^2 \cdot p^2} \cdot e_z \cdot \frac{I_n(pr)}{I_n(p\rho)} - \frac{ih}{p} \cdot h_z \cdot \frac{I'_n(pr)}{I_n(p\rho)} \right) \cdot \gamma$$

$$H_r^{(m)} = \left( \left( 1 - \frac{h^2}{p^2} \right) \cdot \frac{n}{r\omega} \cdot e_z \cdot \frac{I_n(pr)}{I_n(p\rho)} - \frac{ih}{p} \cdot h_z \cdot \frac{I'_n(pr)}{I_n(p\rho)} \right) \cdot \gamma$$

$$H_r^{(m)} = \left( -\frac{n \cdot \varepsilon_m \omega}{r \cdot p^2} \cdot e_z \cdot \frac{I_n(pr)}{I_n(p\rho)} - \frac{ih}{p} \cdot h_z \cdot \frac{I'_n(pr)}{I_n(p\rho)} \right) \cdot \gamma$$

$$H_r^{(h)} = \left( \frac{n}{r\omega} \cdot e_z \cdot \frac{K_n(qr)}{K_n(q\rho)} - \frac{hr}{\omega} \cdot \left( \frac{nh}{r^2 \cdot q^2} \cdot e_z \cdot \frac{K_n(qr)}{K_n(q\rho)} + \frac{i\omega}{r \cdot q} \cdot h_z \cdot \frac{K'_n(qr)}{K_n(q\rho)} \right) \right) \cdot \gamma$$

$$H_r^{(h)} = \left( \frac{n}{r\omega} \cdot e_z \cdot \frac{K_n(qr)}{K_n(q\rho)} - \frac{hr}{\omega} \cdot \frac{nh}{r^2 \cdot q^2} \cdot e_z \cdot \frac{K_n(qr)}{K_n(q\rho)} - \frac{ih}{q} \cdot h_z \cdot \frac{K'_n(qr)}{K_n(q\rho)} \right) \cdot \gamma$$

$$H_r^{(h)} = \left( \left( 1 - \frac{h^2}{q^2} \right) \cdot \frac{n}{r\omega} \cdot e_z \cdot \frac{K_n(qr)}{K_n(q\rho)} - \frac{ih}{q} \cdot h_z \cdot \frac{K'_n(qr)}{K_n(q\rho)} \right) \cdot \gamma$$

$$H_r^{(h)} = \left( -\frac{n \cdot \varepsilon_h \omega}{r \cdot q^2} \cdot e_z \cdot \frac{K_n(qr)}{K_n(q\rho)} - \frac{ih}{q} \cdot h_z \cdot \frac{K'_n(qr)}{K_n(q\rho)} \right) \cdot \gamma$$

Окончательно, находим:

$$E_r^{(m)} = \left( \frac{n \cdot \omega}{r \cdot p^2} \cdot h_z \cdot \frac{I_n(pr)}{I_n(p\rho)} - \frac{ih}{p} \cdot e_z \cdot \frac{I'_n(pr)}{I_n(p\rho)} \right) \cdot \gamma$$

$$E_r^{(h)} = \left( \frac{n\omega}{r \cdot q^2} \cdot h_z \cdot \frac{K_n(qr)}{K_n(q\rho)} - \frac{ih}{q} \cdot e_z \cdot \frac{K'_n(qr)}{K_n(q\rho)} \right) \cdot \gamma \qquad (C.7)$$

$$H_r^{(m)} = \left( -\frac{n \cdot \varepsilon_m \omega}{r \cdot p^2} \cdot e_z \cdot \frac{I_n(pr)}{I_n(p\rho)} - \frac{ih}{p} \cdot h_z \cdot \frac{I'_n(pr)}{I_n(p\rho)} \right) \cdot \gamma$$

$$H_r^{(h)} = \left( -\frac{n \cdot \varepsilon_h \omega}{r \cdot q^2} \cdot e_z \cdot \frac{K_n(qr)}{K_n(q\rho)} - \frac{ih}{q} \cdot h_z \cdot \frac{K'_n(qr)}{K_n(q\rho)} \right) \cdot \gamma$$

## Приложение D.

Рассмотрим компоненты вектора Пойнтинга

$$\vec{S} = \frac{1}{4\pi} \cdot \left[ \vec{E} \times \vec{H} \right]$$

$$\vec{S}_{nk} = \frac{1}{4\pi} \cdot \left( \left[ \vec{E}_{nk} \times \vec{H}_{nk}^* \right] + \left[ \vec{E}_{nk}^* \times \vec{H}_{nk} \right] \right)$$

$$S_\varphi = \frac{2}{4\pi r} \cdot \mathrm{Re}\left( E_z \cdot H_r^* - E_r \cdot H_z^* \right)$$

Как видно из (4.2) и (4.4) при $n = 0$, $S_\varphi = 0$.

$$S_r = \frac{2}{4\pi} \mathrm{Re}\left( rE_\varphi \cdot H_z^* - E_z \cdot rH_\varphi^* \right)$$

$$S_r = \frac{2}{4\pi} \mathrm{Re}\left( rE_\varphi \cdot H_z^* - E_z \cdot rH_\varphi^* \right)$$

из (4.1),(4.2),(4.3) видно, что $E_\varphi, E_z$ — вещественны, а $H_\varphi, H_z$ — мнимы. Значит $S_r = 0$.

## Приложение E1.

Вычисление интеграла (5.6) в металле.

$$\vec{E}_{nk}^{(m)} \cdot \vec{E}_{nk}^{(m)*} = E_{r,nk} \cdot E_{r,nk}^* + r^2 \cdot E_{\varphi,nk} \cdot E_{\varphi,nk}^* + E_{z,nk} \cdot E_{z,nk}^*$$

$$\vec{E}_{nk}^{(m)} \cdot \vec{E}_{nk}^{(m)*} \cdot I_n^2(p\rho) = e_z \cdot e_z^{(m)*} \cdot I_n^2(pr) \cdot \gamma_{nk} \cdot \gamma_{nk}^*$$
$$+ \left( \frac{n\omega}{p^2} \cdot h_z \cdot \frac{I_n(pr)}{r} - \frac{ih}{p} \cdot e_z \cdot I'_n(pr) \right) \cdot \left( \frac{n\omega}{p^2} \cdot h_z^* \cdot \frac{I_n(pr)}{r} + \frac{ih}{p} \cdot e_z^* \cdot I'_n(pr) \right)$$
$$r^2 \cdot \left( \frac{nh}{p^2} \cdot e_z \cdot \frac{I_n(pr)}{r^2} + \frac{i\omega}{p} \cdot h_z \cdot \frac{I'_n(pr)}{r} \right) \cdot \left( \frac{nh}{p^2} \cdot e_z^* \cdot \frac{I_n(pr)}{r^2} - \frac{i\omega}{p} \cdot h_z^* \cdot \frac{I'_n(pr)}{r} \right) \cdot \gamma_{nk} \cdot \gamma_{nk}^*$$

$$\vec{E}_{nk}^{(m)} \cdot \vec{E}_{nk}^{(m)*} \cdot I_n^2(p\rho) \cdot \left(\gamma_{nk} \cdot \gamma_{nk}^*\right)^{-1} = e_z \cdot e_z^* \cdot I_n^2(pr)$$
$$+ e_z \cdot e_z^* \cdot \left(\frac{n^2 h^2}{p^2} \cdot \frac{I_n^2(pr)}{p^2 r^2} + \frac{h^2}{p^2} \cdot I'^2_n(pr)\right) + h_z \cdot h_z^* \cdot \left(\frac{n^2 \omega^2}{p^2} \cdot \frac{I_n^2(pr)}{p^2 r^2} + \frac{\omega^2}{p^2} \cdot I'^2_n(pr)\right) +$$
$$+ \left(h_z \cdot e_z^* - e_z \cdot h_z^*\right) \cdot i \cdot \left(\frac{n\omega}{p^2} \cdot \frac{I_n(pr)}{r} \cdot \frac{h}{p} \cdot I'_n(pr) + \frac{nh}{p^2} \cdot \frac{I_n(pr)}{r} \cdot \frac{\omega}{p} \cdot I'_n(pr)\right)$$

$$\vec{E}_{nk}^{(m)} \cdot \vec{E}_{nk}^{(m)*} \cdot I_n^2(p\rho) \cdot \left(\gamma_{nk} \cdot \gamma_{nk}^*\right)^{-1} = e_z \cdot e_z^* \cdot I_n^2(pr)$$
$$+ \left(e_z \cdot e_z^* \cdot \frac{h^2}{p^2} + h_z \cdot h_z^* \cdot \frac{\omega^2}{p^2}\right) \cdot \left(n^2 \cdot \frac{I_n^2(pr)}{p^2 r^2} + I'^2_n(pr)\right) +$$
$$+ \left(e_z^* \cdot h - e_z \cdot h_z^*\right) \cdot 2i \cdot \left(\frac{nh}{p^2} \cdot \frac{I_n(pr)}{r} \cdot \frac{\omega}{p} \cdot I'_n(pr)\right)$$

$$\vec{H}_{nk}^{(m)} \cdot \vec{H}_{nk}^{(m)} = H_{r,nk} \cdot H_{r,nk}^* + r^2 \cdot H_{\varphi,nk} \cdot H_{\varphi,nk}^* + H_{z,nk} \cdot H_{z,nk}^*$$

$$\vec{H}_{nk}^{(m)} \cdot \vec{H}_{nk}^{(m)} \cdot I_n^2(p\rho) \cdot \left(\gamma_{nk} \cdot \gamma_{nk}^*\right)^{-1} = h_z \cdot I_n^2(pr) \cdot h_z^* +$$
$$\left(-\frac{n \cdot \varepsilon_m \cdot \omega}{p^2} \cdot e_z \cdot \frac{I_n(pr)}{r} - \frac{ih}{p} \cdot h_z \cdot I'_n(pr)\right) \cdot \left(-\frac{n \cdot \varepsilon_m \cdot \omega}{p^2} \cdot e_z^* \cdot \frac{I_n(pr)}{r} + \frac{ih}{p} \cdot h_z^* \cdot I'_n(pr)\right) +$$
$$r^2 \cdot \left(\frac{nh}{p^2} \cdot h_z \cdot \frac{I_n(pr)}{r^2} - \frac{i\omega\varepsilon_m}{p} \cdot e_z \cdot \frac{I'_n(pr)}{r}\right) \cdot \left(\frac{nh}{p^2} \cdot h_z^* \cdot \frac{I_n(pr)}{r^2} + \frac{i\omega\varepsilon_m}{p} \cdot e_z^* \cdot \frac{I'_n(pr)}{r}\right)$$

$$\vec{H}_{nk}^{(m)} \cdot \vec{H}_{nk}^{(m)} \cdot I_n^2(p\rho) \cdot \left(\gamma_{nk} \cdot \gamma_{nk}^*\right)^{-1} = h_z \cdot I_n^2(pr) \cdot h_z^* +$$
$$e_z \cdot e_z^* \cdot \left(\frac{n^2 \cdot \varepsilon_m^2 \cdot \omega^2}{p^2} \cdot \frac{I_n^2(pr)}{p^2 r^2} + \frac{\varepsilon_m^2 \cdot \omega^2}{p^2} \cdot I'^2_n(pr)\right) + h_z \cdot h_z^* \cdot \left(\frac{n^2 \cdot h^2}{p^2} \cdot \frac{I_n^2(pr)}{p^2 r^2} + \frac{h^2}{p^2} \cdot I'^2_n(pr)\right) +$$
$$+ \left(h_z \cdot e_z^* - h_z^* \cdot e_z\right) \cdot i \cdot \left(\frac{nh}{p^2} \cdot \frac{I_n(pr)}{r} \cdot \frac{\omega\varepsilon_m}{p} \cdot I'_n(pr) + \frac{n\omega\varepsilon_m}{p^2} \cdot \frac{I_n(pr)}{r} \cdot \frac{h}{p} \cdot I'_n(pr)\right)$$

$$\vec{H}_{nk}^{(m)} \cdot \vec{H}_{nk}^{(m)} \cdot I_n^2(p\rho) \cdot \left(\gamma_{nk} \cdot \gamma_{nk}^*\right)^{-1} = h_z \cdot h_z^* \cdot I_n^2(pr) +$$
$$\left(e_z \cdot e_z^* \cdot \frac{\varepsilon_m^2 \cdot \omega^2}{p^2} + h_z^{(m)} \cdot h_z^{(m)*} \cdot \frac{h^2}{p^2}\right) \cdot \left(n^2 \cdot \frac{I_n^2(pr)}{p^2 r^2} + I'^2_n(pr)\right) +$$
$$+ \left(h_z \cdot e_z^* - h_z^* \cdot e_z\right) \cdot 2i \cdot \frac{nh}{p^2} \cdot \frac{I_n(pr)}{r} \cdot \frac{\omega\varepsilon_m}{p} \cdot I'_n(pr)$$

$$\left(\varepsilon_m \cdot \vec{E}_{nk}^{(m)} \cdot \vec{E}_{nk}^{(m)*} + \vec{H}_{nk}^{(m)} \cdot \vec{H}_{nk}^{(m)}\right) \cdot I_n^2(p\rho) \cdot \left(\gamma_{nk} \cdot \gamma_{nk}^*\right)^{-1} = \left(\varepsilon_m \cdot e_z \cdot e_z^* + h_z \cdot h_z^*\right) \cdot I_n^2(pr) +$$
$$\left(e_z \cdot e_z^* \cdot \frac{\varepsilon_m^2 \cdot \omega^2}{p^2} + h_z \cdot h_z^* \cdot \frac{h^2}{p^2} + e_z \cdot e_z^* \cdot \frac{\varepsilon_m \cdot h^2}{p^2} + h_z \cdot h_z^* \cdot \frac{\varepsilon_m \cdot \omega^2}{p^2}\right) \cdot \left(n^2 \cdot \frac{I_n^2(pr)}{p^2 r^2} + I'^2_n(pr)\right) +$$
$$+ \left(h_z \cdot e_z^* - h_z^* \cdot e_z\right) \cdot 4i \cdot \frac{nh}{p^2} \cdot \frac{I_n(pr)}{r} \cdot \frac{\omega\varepsilon_m}{p} \cdot I'_n(pr)$$

$$\left(\varepsilon_m \cdot \vec{E}_{nk}^{(m)} \cdot \vec{E}_{nk}^{(m)*} + \vec{H}_{nk}^{(m)} \cdot \vec{H}_{nk}^{(m)}\right) \cdot I_n^2(p\rho) = \left(\varepsilon_m \cdot e_z \cdot e_z^* + h_z \cdot h_z^*\right) \cdot I_n^2(pr) \cdot \gamma_{nk} \cdot \gamma_{nk}^* +$$
$$\left(\varepsilon_m \cdot e_z \cdot e_z^* + h_z \cdot h_z^*\right) \cdot \frac{\varepsilon_m \cdot \omega^2 + h^2}{p^2} \cdot \left(n^2 \cdot \frac{I_n^2(pr)}{p^2 r^2} + I'^2_n(pr)\right) \cdot \gamma_{nk} \cdot \gamma_{nk}^* +$$
$$+ \left(h_z \cdot e_z^* - h_z^* \cdot e_z\right) \cdot 4i \cdot \frac{nh}{p^2} \cdot \frac{\omega \varepsilon_m}{p} \cdot \frac{I_n(pr)}{r} \cdot I'_n(pr) \cdot \gamma_{nk} \cdot \gamma_{nk}^*$$

(E1.1)

Вычислим интегралы

Прямым дифференцированием несложно убедиться, что

$$\int_0^x z \cdot I_n^2(az) \cdot dz = -\frac{x^2}{2}[I'_n(ax)]^2 + \frac{1}{2} \cdot \left(x^2 + \frac{n^2}{a^2}\right) \cdot I_n^2(ax)$$

Используем известные соотношения между бесселевыми функциями:

$$I_{n-1}(z) - I_{n+1}(z) = \frac{2n}{z} \cdot I_n(z)$$

$$I_{n-1}(z) + I_{n+1}(z) = 2 \cdot I'_n(z)$$

$$\int_0^x z \cdot I_n^2(az) \cdot dz = \frac{x^2}{2} \cdot \left(I_n^2(ax) - I_{n-1}(ax) \cdot I_{n+1}(ax)\right)$$

$$F_{I1}^n = \int_0^\rho r \cdot I_n^2(pr) \cdot dr = \frac{1}{p^2} \cdot \int_0^{p\rho} z \cdot I_n^2(z) \cdot dz = \frac{\rho^2}{2} \cdot \left(-I'^2_n(p\rho) + \left(1 + \frac{n^2}{p^2\rho^2}\right) \cdot I_n^2(p\rho)\right)$$

$$F_{I1}^n = \int_0^\rho r \cdot I_n^2(pr) \cdot dr = \frac{1}{p^2} \cdot \int_0^{p\rho} z \cdot I_n^2(z) \cdot dz = \frac{\rho^2}{2} \cdot \left(I_n^2(p\rho) - I_{n-1}(p\rho) \cdot I_{n+1}(p\rho)\right)$$

$$n^2 \cdot \frac{I_n^2(pr)}{p^2 r^2} + I'^2_n(pr) = \frac{1}{2} \cdot \left(I_{n-1}^2(pr) + I_{n+1}^2(pr)\right)$$

$$F_{I2}^n = \int_0^\rho r \cdot \left(n^2 \cdot \frac{I_n^2(pr)}{p^2 r^2} + I'^2_n(pr)\right) \cdot dr = \frac{1}{2p^2} \int_0^{p\rho} z \cdot \left(I_{n-1}^2(z) + I_{n+1}^2(z)\right) \cdot dz$$

$$F_{I2}^n = -\frac{\rho^2}{4}\left([I'_{n-1}(z)]^2 - \left(1 + \frac{(n-1)^2}{z^2}\right) \cdot I_{n-1}^2(z) + [I'_{n+1}(z)]^2 - \left(1 + \frac{(n+1)^2}{z^2}\right) \cdot I_{n+1}^2(z)\right)$$

$$F_{I2}^n = -\frac{\rho^2}{4}\left(\left[\frac{(n-1)}{z}\cdot I_{n-1}(z)+I_n(z)\right]^2 + \left[-\frac{(n+1)}{z}\cdot I_{n+1}(z)+I_n(z)\right]^2\right)$$
$$+\frac{\rho^2}{4}\left(\left(1+\frac{(n-1)^2}{z^2}\right)\cdot I_{n-1}^2(z)+\left(1+\frac{(n+1)^2}{z^2}\right)\cdot I_{n+1}^2(z)\right)$$

$$F_{I2}^n = -\frac{\rho^2}{4}\left(2\cdot\frac{(n-1)}{z}\cdot I_{n-1}(z)\cdot I_n(z)-2\cdot\frac{(n+1)}{z}\cdot I_{n+1}(z)\cdot I_n(z)+2\cdot I_n^2(z)\right)$$
$$+\frac{\rho^2}{4}\left(I_{n-1}^2(z)+I_{n+1}^2(z)\right)$$

$$F_{I2}^n = -\frac{\rho^2}{4}\left(2\cdot\frac{n}{z}\cdot I_{n-1}(z)-\frac{2}{z}\cdot I_{n-1}(z)-2\cdot\frac{n}{z}\cdot I_{n+1}(z)-\frac{2}{z}\cdot I_{n+1}(z)+2\cdot I_n(z)\right)\cdot I_n(z)$$
$$+\frac{\rho^2}{4}\left(I_{n-1}^2(z)+I_{n+1}^2(z)\right)$$

$$F_{I2}^n = -\frac{\rho^2}{4}\left(\frac{4n^2}{z^2}\cdot I_n(z)-\frac{4}{z}\cdot I'_n(z)+2\cdot I_n(z)\right)\cdot I_n(z)$$
$$+\frac{\rho^2}{4}\left(2I'^2_n(z)+\frac{2n^2}{z^2}I_n^2(z)\right)$$

$$F_{I2}^n = -\frac{\rho^2}{2}\left(\frac{n^2}{z^2}I_n^2(z)-\frac{2}{z}\cdot I'_n(z)\cdot I_n(z)+I_n^2(z)-I'^2_n(z)\right)$$

$$F_{I3}^n = \int_0^\rho r\cdot\left(\frac{I_n(pr)}{r}\cdot I'_n(pr)\right)\cdot dr = \frac{1}{p}\int_0^{p\rho} I_n(z)\cdot I'_n(z)\cdot dz = \frac{1}{2p}I_n^2(p\rho)$$

========-==========-==============-=

Умножим обе части (E1.1) на $r$ и проинтегрируем от 0 до $\rho$.

Из (5.6) видим, что

$$\frac{4}{L}\cdot W_{nk}^{(m)}\cdot I_n^2(p\rho) = \left(\varepsilon_m\cdot e_z\cdot e_z^* + h_z\cdot h_z^*\right)\cdot F_{I1}^n\cdot\gamma_{nk}\cdot\gamma_{nk}^* +$$
$$\left(\varepsilon_m\cdot e_z\cdot e_z^* + h_z\cdot h_z^*\right)\cdot\frac{\varepsilon_m\cdot\omega^2+h^2}{p^2}\cdot F_{I2}^n\cdot\gamma_{nk}\cdot\gamma_{nk}^* +$$
$$+\left(h_z\cdot e_z^* - h_z^*\cdot e_z\right)\cdot 4i\cdot\frac{nh}{p^2}\cdot\frac{\omega\varepsilon_m}{p}\cdot F_{I3}^n\cdot\gamma_{nk}\cdot\gamma_{nk}^*$$

$$\frac{4}{L}\cdot W_{nk}^{(m)}\cdot I_n^2(p\rho)\cdot\left(\gamma_{nk}\cdot\gamma_{nk}^*\right)^{-1} = \left(\varepsilon_m\cdot e_z\cdot e_z^* + h_z\cdot h_z^*\right)\cdot F_{I1}^n +$$
$$\left(\varepsilon_m\cdot e_z\cdot e_z^* + h_z\cdot h_z^*\right)\cdot\frac{\varepsilon_m\cdot\omega^2+h^2}{p^2}\cdot F_{I2}^n +$$
$$+\left(h_z\cdot e_z^* - h_z^*\cdot e_z\right)\cdot 4i\cdot\frac{nh}{p^2}\cdot\frac{\omega\varepsilon_m}{p}\cdot F_{I3}^n$$

$$\frac{4}{L} \cdot W_{nk}^{(m)} \cdot I_n^2(p\rho) \cdot \left(\gamma_{nk} \cdot \gamma_{nk}^*\right)^{-1} = \left(\varepsilon_m \cdot e_z \cdot e_z^* + h_z \cdot h_z^*\right) \cdot \frac{\rho^2}{2} \cdot \left(-I_n'^2(p\rho) + \left(1 + \frac{n^2}{p^2\rho^2}\right) \cdot I_n^2(p\rho)\right) -$$

$$- \left(\varepsilon_m \cdot e_z \cdot e_z^* + h_z \cdot h_z^*\right) \cdot \frac{\varepsilon_m \cdot \omega^2 + h^2}{p^2} \cdot \frac{\rho^2}{2} \left(\frac{n^2}{p^2\rho^2} I_n^2(p\rho) - \frac{2}{p\rho} \cdot I_n'(p\rho) \cdot I_n(p\rho) + I_n^2(p\rho) - I_n'^2(p\rho)\right) +$$

$$+ \left(h_z \cdot e_z^* - h_z^* \cdot e_z\right) \cdot 4i \cdot \frac{nh}{p^2} \cdot \frac{\omega\varepsilon_m}{p} \cdot \frac{1}{2p} I_n^2(p\rho)$$

$$\frac{4}{L} \cdot W_{nk}^{(m)} \cdot I_n^2(p\rho) \cdot \left(\gamma_{nk} \cdot \gamma_{nk}^*\right)^{-1} = \left(\varepsilon_m \cdot e_z \cdot e_z^* + h_z \cdot h_z^*\right) \cdot \frac{\rho^2}{2} \cdot \left(-I_n'^2(p\rho) + \left(1 + \frac{n^2}{p^2\rho^2}\right) \cdot I_n^2(p\rho)\right) -$$

$$- \left(\varepsilon_m \cdot e_z \cdot e_z^* + h_z \cdot h_z^*\right) \cdot \frac{\varepsilon_m \cdot \omega^2 + h^2}{p^2} \cdot \frac{\rho^2}{2} \left(\frac{n^2}{p^2\rho^2} I_n^2(p\rho) + I_n^2(p\rho) - I_n'^2(p\rho)\right) +$$

$$+ \left(\varepsilon_m \cdot e_z \cdot e_z^* + h_z \cdot h_z^*\right) \cdot \frac{\varepsilon_m \cdot \omega^2 + h^2}{p^2} \cdot \frac{\rho^2}{2} \cdot \frac{2}{p\rho} \cdot I_n'(p\rho) \cdot I_n(p\rho) + \left(h_z \cdot e_z^* - h_z^* \cdot e_z\right) \cdot 2i \cdot \frac{nh}{p^2} \cdot \frac{\omega\varepsilon_m}{p^2} \cdot I_n^2(p\rho)$$

$$\frac{4}{L} \cdot W_{nk}^{(m)} \cdot I_n^2(p\rho) \cdot \left(\gamma_{nk} \cdot \gamma_{nk}^*\right)^{-1} =$$

$$- \left(\varepsilon_m \cdot e_z \cdot e_z^* + h_z \cdot h_z^*\right) \cdot \left(\frac{\varepsilon_m \cdot \omega^2 + h^2}{p^2} - 1\right) \cdot \frac{\rho^2}{2} \left(\frac{n^2}{p^2\rho^2} I_n^2(p\rho) + I_n^2(p\rho) - I_n'^2(p\rho)\right) +$$

$$+ \left(\varepsilon_m \cdot e_z \cdot e_z^* + h_z \cdot h_z^*\right) \cdot \frac{\varepsilon_m \cdot \omega^2 + h^2}{p^3} \cdot \rho \cdot I_n'(p\rho) \cdot I_n(p\rho) + \left(h_z \cdot e_z^* - h_z^* \cdot e_z\right) \cdot 2i \cdot \frac{nh}{p^2} \cdot \frac{\omega\varepsilon_m}{p^2} \cdot I_n^2(p\rho)$$

$$\frac{4}{L} \cdot W_{nk}^{(m)} \cdot I_n^2(p\rho) \cdot \left(\gamma_{nk} \cdot \gamma_{nk}^*\right)^{-1} =$$

$$- \left(\varepsilon_m \cdot e_z \cdot e_z^* + h_z \cdot h_z^*\right) \cdot \frac{\varepsilon_m \cdot \omega^2 \cdot \rho^2}{p^2} \left(\frac{n^2}{p^2\rho^2} I_n^2(p\rho) + I_n^2(p\rho) - I_n'^2(p\rho)\right) +$$

$$+ \left(\varepsilon_m \cdot e_z \cdot e_z^* + h_z \cdot h_z^*\right) \cdot \frac{\varepsilon_m \cdot \omega^2 + h^2}{p^3} \cdot \rho \cdot I_n'(p\rho) \cdot I_n(p\rho) + \left(h_z \cdot e_z^* - h_z^* \cdot e_z\right) \cdot 2i \cdot \frac{nh}{p^2} \cdot \frac{\omega\varepsilon_m}{p^2} \cdot I_n^2(p\rho)$$

$$\frac{4}{L} \cdot W_{nk}^{(m)} \cdot \left(\gamma_{nk} \cdot \gamma_{nk}^*\right)^{-1} = -\left(\varepsilon_m \cdot e_z \cdot e_z^* + h_z \cdot h_z^*\right) \cdot \frac{\varepsilon_m \cdot \omega^2 \cdot \rho^2}{p^2} \left(\frac{n^2}{p^2\rho^2} + 1 - \frac{I_n'^2(p\rho)}{I_n^2(p\rho)}\right) +$$

$$+ \left(\varepsilon_m \cdot e_z \cdot e_z^* + h_z \cdot h_z^*\right) \cdot \frac{\varepsilon_m \cdot \omega^2 + h^2}{p^3} \cdot \rho \cdot \frac{I_n'(p\rho)}{I_n(p\rho)} + \left(h_z \cdot e_z^* - h_z^* \cdot e_z\right) \cdot 2i \cdot \frac{nh}{p^2} \cdot \frac{\omega\varepsilon_m}{p^2}$$

$$\frac{4}{L} \cdot W_{nk}^{(m)} \cdot \left(\gamma_{nk} \cdot \gamma_{nk}^*\right)^{-1} = -\left(\varepsilon_m \cdot e_z \cdot e_z^* + h_z \cdot h_z^*\right) \cdot \frac{\varepsilon_m \cdot \omega^2}{p^4} \left(n^2 + p^2\rho^2 - D_I^2\right) +$$

$$+ \left(\varepsilon_m \cdot e_z \cdot e_z^* + h_z \cdot h_z^*\right) \cdot \frac{\varepsilon_m \cdot \omega^2 + h^2}{p^4} \cdot D_I + \left(h_z \cdot e_z^* - h_z^* \cdot e_z\right) \cdot 2i \cdot \frac{nh}{p^2} \cdot \frac{\omega\varepsilon_m}{p^2}$$

$$\frac{4}{L} \cdot W_{nk}^{(m)} \cdot \left(\gamma_{nk} \cdot \gamma_{nk}^*\right)^{-1} = -\left(\varepsilon_m \cdot e_z \cdot e_z^* + h_z \cdot h_z^*\right) \cdot \frac{\varepsilon_m \cdot \omega^2}{p^4}\left(n^2 + p^2\rho^2 - D_I^2\right) +$$

$$+\left(\varepsilon_m \cdot e_z \cdot e_z^* + h_z \cdot h_z^*\right) \cdot \frac{2 \cdot \varepsilon_m \cdot \omega^2 + p^2}{p^4} \cdot D_I + \left(h_z \cdot e_z^* - h_z^* \cdot e_z\right) \cdot 2i \cdot \frac{nh}{p^2} \cdot \frac{\omega \varepsilon_m}{p^2}$$

$$\frac{4}{L} \cdot W_{nk}^{(m)} \cdot \left(\gamma_{nk} \cdot \gamma_{nk}^*\right)^{-1} = -\left(\varepsilon_m \cdot e_z \cdot e_z^* + h_z \cdot h_z^*\right) \cdot \frac{\varepsilon_m \cdot \omega^2}{p^4}\left(n^2 + p^2\rho^2 - D_I^2 - 2 \cdot D_I\right) +$$

$$+\left(\varepsilon_m \cdot e_z \cdot e_z^* + h_z \cdot h_z^*\right) \cdot \frac{1}{p^2} \cdot D_I + \left(h_z \cdot e_z^* - h_z^* \cdot e_z\right) \cdot 2i \cdot \frac{nh}{p^2} \cdot \frac{\omega \varepsilon_m}{p^2}$$

$$W_{nk}^{(m)} = -\left(\varepsilon_m \cdot e_z \cdot e_z^* + h_z \cdot h_z^*\right) \cdot \frac{L}{4} \cdot \frac{\varepsilon_m \cdot \omega^2}{p^4}\left(n^2 + p^2\rho^2 - D_I^2 - 2 \cdot D_I\right) \cdot \gamma_{nk} \cdot \gamma_{nk}^* +$$

$$+\frac{L}{4} \cdot \left(\varepsilon_m \cdot e_z \cdot e_z^* + h_z \cdot h_z^*\right) \cdot \frac{1}{p^2} \cdot D_I \cdot \gamma_{nk} \cdot \gamma_{nk}^* + \frac{L}{4} \cdot \left(h_z \cdot e_z^* - h_z^* \cdot e_z\right) \cdot 2i \cdot \frac{nh}{p^2} \cdot \frac{\omega \varepsilon_m}{p^2} \cdot \gamma_{nk} \cdot \gamma_{nk}^*$$

$$W_{nk}^{(m)} = -\frac{L}{4} \cdot \left(\varepsilon_m \cdot e_z \cdot e_z^* + h_z \cdot h_z^*\right) \cdot \frac{\varepsilon_m \cdot \omega^2 \cdot \rho^2}{p^2} \cdot \left(1 + \frac{n^2 - D_I^2 - 2 \cdot D_I}{p^2 \cdot \rho^2}\right) \cdot \gamma_{nk} \cdot \gamma_{nk}^* +$$

$$+\frac{L}{4} \cdot \left(\varepsilon_m \cdot e_z \cdot e_z^* + h_z \cdot h_z^*\right) \cdot \frac{1}{p^2} \cdot D_I \cdot \gamma_{nk} \cdot \gamma_{nk}^* + \frac{L}{2} \cdot \left(h_z \cdot e_z^* - h_z^* \cdot e_z\right) \cdot \frac{inh \cdot \varepsilon_m \omega}{p^4} \cdot \gamma_{nk} \cdot \gamma_{nk}^*$$

**Приложение Е2.**

Вычисление интеграла (5.6) в диэлектрике.

$$\vec{E}_{nk} \cdot \vec{E}_{nk}^* = E_{r,nk} \cdot E_{r,nk}^* + r^2 \cdot E_{\varphi,nk} \cdot E_{\varphi,nk}^* + E_{z,nk} \cdot E_{z,nk}^*$$

$$\vec{E}_{nk}^{(h)} \cdot \vec{E}_{nk}^{(h)*} \cdot K_n^2(q\rho) \cdot \left(\gamma_{nk} \cdot \gamma_{nk}^*\right)^{-1} = e_z \cdot e_z^* \cdot K_n^2(qr) +$$

$$+\left(\frac{n\omega}{q^2} \cdot h_z \cdot \frac{K_n(qr)}{r} - \frac{ih}{q} \cdot e_z \cdot K'_n(qr)\right)\left(\frac{n\omega}{q^2} \cdot h_z^* \cdot \frac{K_n(qr)}{r} + \frac{ih}{q} \cdot e_z^* \cdot K'_n(qr)\right) +$$

$$+ r^2 \cdot \left(\frac{nh}{r^2 \cdot q^2} \cdot e_z \cdot K_n(qr) + \frac{i\omega}{r \cdot q} h_z \cdot K'_n(qr)\right) \cdot \left(\frac{nh}{r^2 \cdot q^2} \cdot e_z^* \cdot K_n(qr) - \frac{i\omega}{r \cdot q} h_z^* \cdot K'_n(qr)\right)$$

$$\vec{E}_{nk}^{(h)} \cdot \vec{E}_{nk}^{(h)*} \cdot K_n^2(q\rho) \cdot \left(\gamma_{nk} \cdot \gamma_{nk}^*\right)^{-1} = e_z^{(h)} \cdot e_z^{(h)*} \cdot K_n^2(qr) +$$

$$+\left(h_z \cdot h_z^* \cdot \left(\frac{n^2\omega^2}{q^4} \cdot \frac{K_n^2(qr)}{r^2} + \frac{\omega^2}{q^2} \cdot (K'_n(qr))^2\right) + e_z \cdot e_z^* \cdot \left(\frac{n^2h^2}{q^4} \cdot \frac{K_n^2(qr)}{r^2} + \frac{h^2}{q^2} \cdot (K'_n(qr))^2\right)\right) +$$

$$+ h_z \cdot e_z^* \cdot \left(\frac{\omega}{q^2}\frac{n \cdot ih}{q} \cdot \frac{K_n(qr)}{r} \cdot K'_n(qr) + \frac{i\omega}{q} \cdot K'_n(qr) \cdot \frac{K_n(qr)}{r} \cdot \frac{nh}{q^2}\right)$$

$$+ h_z^* \cdot e_z \cdot \left(-\frac{\omega}{q^2}\frac{n \cdot ih}{q} \cdot \frac{K_n(qr)}{r} \cdot K'_n(qr) - \frac{i\omega}{q} \cdot K'_n(qr) \cdot \frac{K_n(qr)}{r} \cdot \frac{nh}{q^2}\right)$$

$$\vec{E}_{nk}^{(h)} \cdot \vec{E}_{nk}^{(h)*} \cdot K_n^2(q\rho) \cdot (\gamma_{nk} \cdot \gamma_{nk}^*)^{-1} = e_z \cdot e_z^* \cdot K_n^2(qr) +$$
$$+ \left( h_z \cdot h_z^* \cdot \frac{\omega^2}{q^2} \cdot \left( n^2 \cdot \frac{K_n^2(qr)}{q^2 \cdot r^2} + (K'_n(qr))^2 \right) + e_z \cdot e_z^* \cdot \frac{h^2}{q^2} \cdot \left( n^2 \cdot \frac{K_n^2(qr)}{q^2 \cdot r^2} + (K'_n(qr))^2 \right) \right) +$$
$$+ 2i \cdot \left( h_z^{(h)} \cdot e_z^{(h)*} - h_z^{(h)*} \cdot e_z^{(h)} \right) \cdot \left( \frac{\omega}{q^2} \frac{n \cdot h}{q} \cdot \frac{K_n(qr)}{r} \cdot K'_n(qr) \right)$$

$$\vec{E}_{nk}^{(h)} \cdot \vec{E}_{nk}^{(h)*} \cdot K_n^2(q\rho) \cdot (\gamma_{nk} \cdot \gamma_{nk}^*)^{-1} = e_z \cdot e_z^* \cdot K_n^2(qr) +$$
$$+ \left( h_z \cdot h_z^* \cdot \frac{\omega^2}{q^2} + e_z \cdot e_z^* \cdot \frac{h^2}{q^2} \right) \cdot \left( n^2 \cdot \frac{K_n^2(qr)}{q^2 \cdot r^2} + (K'_n(qr))^2 \right) +$$
$$+ 2i \cdot \left( h_z \cdot e_z^* - h_z^* \cdot e_z \right) \cdot \frac{\omega}{q^2} \frac{n \cdot h}{q} \cdot \frac{K_n(qr)}{r} \cdot K'_n(qr)$$

$$\vec{H}_{nk}^{(h)} \cdot \vec{H}_{nk}^{(h)} = H_{r,nk} \cdot H_{r,nk}^* + r^2 \cdot H_{\varphi,nk} \cdot H_{\varphi,nk}^* + H_{z,nk} \cdot H_{z,nk}^*$$

$$\vec{H}_{nk}^{(h)} \cdot \vec{H}_{nk}^{(h)} \cdot K_n^2(q\rho) \cdot (\gamma_{nk} \cdot \gamma_{nk}^*)^{-1} = h_z^{(h)} \cdot h_z^{(h)*} \cdot K_n^2(qr) +$$
$$+ \left( -\frac{n \cdot \varepsilon_h \omega}{q^2} \cdot e_z \cdot \frac{K_n(qr)}{r} - \frac{ih}{q} h_z \cdot K'_n(qr) \right) \cdot \left( -\frac{n \cdot \varepsilon_h \omega}{q^2} \cdot e_z^* \cdot \frac{K_n(qr)}{r} + \frac{ih}{q} h_z^* \cdot K'_n(qr) \right) +$$
$$+ r^2 \cdot \left( \frac{nh}{r^2 \cdot q^2} \cdot h_z \cdot K_n(qr) - \frac{i\omega \varepsilon_h}{r \cdot q} \cdot e_z \cdot K'_n(qr) \right) \cdot \left( \frac{nh}{r^2 \cdot q^2} \cdot h_z^* \cdot K_n(qr) + \frac{i\omega \varepsilon_h}{r \cdot q} \cdot e_z^* \cdot K'_n(qr) \right)$$

$$\vec{H}_{nk}^{(h)} \cdot \vec{H}_{nk}^{(h)} \cdot K_n^2(q\rho) \cdot (\gamma_{nk} \cdot \gamma_{nk}^*)^{-1} = h_z \cdot h_z^* \cdot K_n^2(qr) +$$
$$+ \left( e_z \cdot e_z^* \cdot \left( \left( \frac{n \cdot \varepsilon_h \omega}{q^2} \cdot \frac{K_n(qr)}{r} \right)^2 + \left( \frac{\omega \varepsilon_h}{q} \cdot K'_n(qr) \right)^2 \right) + h_z \cdot h_z^* \cdot \left( \frac{n^2 h^2}{q^2} \cdot \frac{K_n^2(qr)}{r^2 q^2} + \frac{h^2}{q^2} \cdot (K'_n(qr))^2 \right) \right) +$$
$$+ e_z \cdot h_z^* \cdot \left( -\frac{n \cdot \varepsilon_h \omega}{q^2} \cdot \frac{K_n(qr)}{r} \cdot \frac{ih}{q} \cdot K'_n(qr) - \frac{i\omega \varepsilon_h}{q} \cdot K'_n(qr) \cdot \frac{nh}{q^2} \cdot \frac{K_n(qr)}{r} \right)$$
$$+ e_z^* \cdot h_z \cdot \left( \frac{n \cdot \varepsilon_h \omega}{q^2} \cdot \frac{K_n(qr)}{r} \cdot \frac{ih}{q} \cdot K'_n(qr) + \frac{i\omega \varepsilon_h}{q} \cdot K'_n(qr) \cdot \frac{nh}{q^2} \cdot \frac{K_n(qr)}{r} \right)$$

$$\vec{H}_{nk}^{(h)} \cdot \vec{H}_{nk}^{(h)} \cdot K_n^2(q\rho) \cdot (\gamma_{nk} \cdot \gamma_{nk}^*)^{-1} = h_z \cdot h_z^* \cdot K_n^2(qr) +$$
$$+ \left( e_z \cdot e_z^* \cdot \left( \frac{\omega \varepsilon_h}{q} \right)^2 \cdot \left( \left( n \cdot \frac{K_n(qr)}{qr} \right)^2 + (K'_n(qr))^2 \right) + h_z \cdot h_z^* \cdot \frac{h^2}{q^2} \cdot \left( n^2 \cdot \frac{K_n^2(qr)}{r^2 q^2} + (K'_n(qr))^2 \right) \right) +$$
$$+ 2i \cdot \left( e_z^* \cdot h_z - e_z \cdot h_z^* \right) \cdot \left( \frac{n \cdot \varepsilon_h \omega}{q^2} \cdot \frac{K_n(qr)}{r} \cdot \frac{h}{q} \cdot K'_n(qr) \right)$$

$$\vec{H}_{nk}^{(h)} \cdot \vec{H}_{nk}^{(h)} \cdot K_n^2(q\rho) \cdot (\gamma_{nk} \cdot \gamma_{nk}^*)^{-1} = h_z \cdot h_z^* \cdot K_n^2(qr) +$$
$$+ \left( e_z \cdot e_z^* \cdot \left( \frac{\omega \varepsilon_h}{q} \right)^2 + h_z \cdot h_z^* \cdot \frac{h^2}{q^2} \right) \cdot \left( n^2 \cdot \frac{K_n^2(qr)}{r^2 q^2} + (K'_n(qr))^2 \right) +$$
$$+ 2i \cdot \left( e_z^* \cdot h_z - e_z \cdot h_z^* \right) \cdot \left( \frac{n \cdot \varepsilon_h \omega}{q^2} \cdot \frac{K_n(qr)}{r} \cdot \frac{h}{q} \cdot K'_n(qr) \right)$$

$$\left(\varepsilon_h \cdot \vec{E}_{nk}^{(h)} \cdot \vec{E}_{nk}^{(h)*} + \vec{H}_{nk}^{(h)} \cdot \vec{H}_{nk}^{(h)*}\right) \cdot K_n^2(q\rho) \cdot \left(\gamma_{nk} \cdot \gamma_{nk}^*\right)^{-1} = \left(\varepsilon_h \cdot e_z \cdot e_z^* + h_z \cdot h_z^*\right) \cdot K_n^2(qr) +$$
$$+ \left(e_z \cdot e_z^* \cdot \left(\varepsilon_h \cdot \frac{h^2}{q^2} + \left(\frac{\omega \varepsilon_h}{q}\right)^2\right) + h_z \cdot h_z^* \cdot \left(\varepsilon_h \cdot \frac{\omega^2}{q^2} + \frac{h^2}{q^2}\right)\right) \cdot \left(n^2 \cdot \frac{K_n^2(qr)}{r^2 q^2} + (K'_n(qr))^2\right) +$$
$$+ 4i \cdot \left(e_z^* \cdot h_z - e_z \cdot h_z^*\right) \cdot \left(\frac{n \cdot \varepsilon_h \omega}{q^2} \cdot \frac{K_n(qr)}{r} \cdot \frac{h}{q} \cdot K'_n(qr)\right)$$

| |
|---|
| $\left(\varepsilon_h \cdot \vec{E}_{nk}^{(h)} \cdot \vec{E}_{nk}^{(h)*} + \vec{H}_{nk}^{(h)} \cdot \vec{H}_{nk}^{(h)*}\right) \cdot K_n^2(q\rho) \cdot \left(\gamma_{nk} \cdot \gamma_{nk}^*\right)^{-1} = \left(\varepsilon_h \cdot e_z \cdot e_z^* + h_z \cdot h_z^*\right) \cdot K_n^2(qr) +$ <br> $+ \left(\varepsilon_h \cdot e_z \cdot e_z^* + h_z \cdot h_z^*\right) \cdot \left(\varepsilon_h \cdot \frac{\omega^2}{q^2} + \frac{h^2}{q^2}\right) \cdot \left(n^2 \cdot \frac{K_n^2(qr)}{r^2 q^2} + (K'_n(qr))^2\right) +$ <br> $+ 4i \cdot \left(e_z^* \cdot h_z - e_z \cdot h_z^*\right) \cdot \left(\frac{n \cdot \varepsilon_h \omega}{q^2} \cdot \frac{K_n(qr)}{r} \cdot \frac{h}{q} \cdot K'_n(qr)\right)$     (Е2.1) |

Вычислим интегралы

$$\int_x^\infty z \cdot K_n^2(az) \cdot dz = \frac{x^2}{2}[K'_n(ax)]^2 - \frac{1}{2} \cdot \left(x^2 + \frac{n^2}{a^2}\right) \cdot K_n^2(ax)$$

$$F_{K1}^n = \int_\rho^\infty r \cdot K_n^2(pr) \cdot dr = \frac{1}{p^2} \cdot \int_{p\rho}^\infty z \cdot K_n^2(z) \cdot dz = \frac{\rho^2}{2} \cdot \left((K'_n(p\rho))^2 - \left(1 + \frac{n^2}{p^2\rho^2}\right) \cdot K_n^2(p\rho)\right)$$

$$F_{K2}^n = \int_\rho^\infty r \cdot \left(n^2 \cdot \frac{K_n^2(pr)}{p^2 r^2} + K'^2_n(pr)\right) \cdot dr = \frac{1}{2p^2} \int_{p\rho}^\infty z \cdot \left(K_{n-1}^2(z) + K_{n+1}^2(z)\right) \cdot dz$$

$$z = p\rho$$

$$F_{K2}^n = \frac{\rho^2}{4} \cdot \left((K'_{n-1}(z))^2 - \left(1 + \frac{(n-1)^2}{z^2}\right) \cdot K_{n-1}^2(z) + (K'_{n+1}(z))^2 - \left(1 + \frac{(n+1)^2}{z^2}\right) \cdot K_{n+1}^2(z)\right)$$

Воспользуемся рекуррентными соотношениями:

$$K_{n-1}(z) - K_{n+1}(z) = -\frac{2n}{z} \cdot K_n(z)$$

$$K_{n-1}(z) + K_{n+1}(z) = -2 \cdot K'_n(z)$$

$$F_{K2}^n = \frac{\rho^2}{4} \cdot \left(\left(K_n(z) - \frac{n-1}{z} \cdot K_{n-1}(z)\right)^2 + \left(K_n(z) + \frac{n+1}{z} \cdot K_{n+1}(z)\right)^2\right)$$
$$- \frac{\rho^2}{4} \cdot \left(\left(1 + \frac{(n-1)^2}{z^2}\right) \cdot K_{n-1}^2(z) + \left(1 + \frac{(n+1)^2}{z^2}\right) \cdot K_{n+1}^2(z)\right)$$

$$F_{K2}^n = \frac{\rho^2}{4} \cdot \left( K_n^2(z) - 2K_n(z) \cdot \frac{n-1}{z} \cdot K_{n-1}(z) + K_n^2(z) + 2 \cdot K_n(z) \cdot \frac{n+1}{z} \cdot K_{n+1}(z) \right)$$
$$- \frac{\rho^2}{4} \cdot \left( K_{n-1}^2(z) + K_{n+1}^2(z) \right)$$

$$F_{K2}^n = \frac{\rho^2}{2} \cdot \left( -\frac{n-1}{z} \cdot K_{n-1}(z) + K_n(z) + \frac{n+1}{z} \cdot K_{n+1}(z) \right) \cdot K_n(z)$$
$$- \frac{\rho^2}{4} \cdot \left( 2K'^2_n(z) + \frac{2n^2}{z^2} K_n^2(z) \right)$$

$$F_{K2}^n = \frac{\rho^2}{2} \cdot \left( -\frac{n}{z} \cdot K_{n-1}(z) + \frac{n}{z} \cdot K_{n+1}(z) + K_n(z) + \frac{1}{z} \cdot K_{n-1}(z) + \frac{1}{z} \cdot K_{n+1}(z) \right) \cdot K_n(z)$$
$$- \frac{\rho^2}{2} \cdot \left( K'^2_n(z) + \frac{n^2}{z^2} K_n^2(z) \right)$$

$$F_{K2}^n = \frac{\rho^2}{2} \cdot \left( \frac{2n^2}{z^2} \cdot K_n(z) + K_n(z) - \frac{2}{z} \cdot K'_n(z) \right) \cdot K_n(z)$$
$$- \frac{\rho^2}{2} \cdot \left( K'^2_n(z) + \frac{n^2}{z^2} K_n^2(z) \right)$$

$$F_{K2}^n = \frac{\rho^2}{2} \cdot \left( \left( 1 + \frac{n^2}{z^2} \right) \cdot K_n^2(z) - \frac{2}{z} \cdot K'_n(z) \cdot K_n(z) - K'^2_n(z) \right)$$

$$F_{K3}^n = \int_\rho^\infty r \cdot \frac{K_n(qr)}{r} K'_n(qr) \cdot dr = \frac{1}{q^2} \int_{q\rho}^\infty K_n(z) \cdot K'_n(z) \cdot dz = \frac{1}{2q^2} K_n^2(q\rho)$$

Из (5.6) и (E2.1) видим, что

$$\frac{4}{L} \cdot W_{nk}^{(h)} \cdot K_n^2(q\rho) \cdot (\gamma_{nk} \cdot \gamma_{nk}^*)^{-1} = (\varepsilon_h \cdot e_z \cdot e_z^* + h_z \cdot h_z^*) \cdot F_{K1}^n +$$
$$+ (\varepsilon_h \cdot e_z \cdot e_z^* + h_z \cdot h_z^*) \cdot \left( \varepsilon_h \frac{\omega^2}{q^2} + \frac{h^2}{q^2} \right) \cdot F_{K2}^n +$$
$$+ 4i \cdot (e_z^* \cdot h_z - e_z \cdot h_z^*) \cdot F_{K3}^n$$

$$\frac{4}{L} \cdot W_{nk}^{(h)} \cdot K_n^2(q\rho) \cdot (\gamma_{nk} \cdot \gamma_{nk}^*)^{-1} = (\varepsilon_h \cdot e_z \cdot e_z^* + h_z \cdot h_z^*) \cdot \frac{\rho^2}{2} \cdot \left( (K'_n(p\rho))^2 - \left( 1 + \frac{n^2}{p^2\rho^2} \right) \cdot K_n^2(p\rho) \right) +$$
$$(\varepsilon_h \cdot e_z \cdot e_z^* + h_z \cdot h_z^*) \cdot \left( \varepsilon_h \frac{\omega^2}{q^2} + \frac{h^2}{q^2} \right) \cdot \frac{\rho^2}{2} \cdot \left( \left( 1 + \frac{n^2}{q^2\rho^2} \right) \cdot K_n^2(q\rho) - \frac{2}{q\rho} \cdot K'_n(q\rho) \cdot K_n(q\rho) - K'^2_n(q\rho) \right)$$
$$+ 4i \cdot (e_z^* \cdot h_z - e_z \cdot h_z^*) \cdot \frac{nh \cdot \varepsilon_h \omega}{q^2} \cdot \frac{1}{2q^2} K_n^2(q\rho)$$

$$\frac{4}{L} \cdot W_{nk}^{(h)} \cdot K_n^2(q\rho) \cdot (\gamma_{nk} \cdot \gamma_{nk}^*)^{-1} = (\varepsilon_h \cdot e_z \cdot e_z^* + h_z \cdot h_z^*) \cdot \frac{\rho^2}{2} \cdot \left( (K'_n(p\rho))^2 - \left(1 + \frac{n^2}{p^2\rho^2}\right) \cdot K_n^2(p\rho) \right) +$$

$$(\varepsilon_h \cdot e_z \cdot e_z^* + h_z \cdot h_z^*) \cdot \left( \varepsilon_h \cdot \frac{\omega^2}{q^2} + \frac{h^2}{q^2} \right) \cdot \frac{\rho^2}{2} \cdot \left( \left(1 + \frac{n^2}{q^2\rho^2}\right) \cdot K_n^2(q\rho) - K'^2_n(q\rho) \right)$$

$$- (\varepsilon_h \cdot e_z \cdot e_z^* + h_z \cdot h_z^*) \cdot \left( \varepsilon_h \cdot \frac{\omega^2}{q^2} + \frac{h^2}{q^2} \right) \cdot \frac{\rho^2}{2} \cdot \frac{2}{q\rho} \cdot K'_n(q\rho) \cdot K_n(q\rho)$$

$$+ 4i \cdot (e_z^* \cdot h_z - e_z \cdot h_z^*) \cdot \frac{nh \cdot \varepsilon_h \omega}{q^2} \cdot \frac{1}{2q^2} K_n^2(q\rho)$$

$$\frac{4}{L} \cdot W_{nk}^{(h)} \cdot K_n^2(q\rho) \cdot (\gamma_{nk} \cdot \gamma_{nk}^*)^{-1} = +4i \cdot (e_z^* \cdot h_z - e_z \cdot h_z^*) \cdot \frac{nh \cdot \varepsilon_h \omega}{q^2} \cdot \frac{1}{2q^2} K_n^2(q\rho) +$$

$$+ (\varepsilon_h \cdot e_z \cdot e_z^* + h_z \cdot h_z^*) \cdot \left( \varepsilon_h \cdot \frac{\omega^2}{q^2} + \frac{h^2}{q^2} - 1 \right) \cdot \frac{\rho^2}{2} \cdot \left( \left(1 + \frac{n^2}{q^2\rho^2}\right) \cdot K_n^2(q\rho) - K'^2_n(q\rho) \right)$$

$$- (\varepsilon_h \cdot e_z \cdot e_z^* + h_z \cdot h_z^*) \cdot \left( \varepsilon_h \cdot \frac{\omega^2}{q^2} + \frac{h^2}{q^2} \right) \cdot \frac{\rho^2}{2} \cdot \frac{2}{q\rho} \cdot K'_n(q\rho) \cdot K_n(q\rho)$$

$$\frac{4}{L} \cdot W_{nk}^{(h)} \cdot K_n^2(q\rho) \cdot (\gamma_{nk} \cdot \gamma_{nk}^*)^{-1} = +\frac{2i}{q^2} \cdot (e_z^* \cdot h_z - e_z \cdot h_z^*) \cdot \frac{nh \cdot \varepsilon_h \omega}{q^2} \cdot K_n^2(q\rho) +$$

$$+ (\varepsilon_h \cdot e_z \cdot e_z^* + h_z \cdot h_z^*) \cdot \frac{\varepsilon_h \cdot \omega^2}{q^2} \cdot \rho^2 \cdot \left( \left(1 + \frac{n^2}{q^2\rho^2}\right) \cdot K_n^2(q\rho) - K'^2_n(q\rho) \right)$$

$$- (\varepsilon_h \cdot e_z \cdot e_z^* + h_z \cdot h_z^*) \cdot \frac{\varepsilon_h \cdot \omega^2 + h^2}{q^3} \cdot \rho \cdot K'_n(q\rho) \cdot K_n(q\rho)$$

$$\frac{4}{L} \cdot W_{nk}^{(h)} \cdot (\gamma_{nk} \cdot \gamma_{nk}^*)^{-1} = +\frac{2i}{q^2} \cdot \frac{nh \cdot \varepsilon_h \omega}{q^2} \cdot (e_z^* \cdot h_z - e_z \cdot h_z^*) +$$

$$+ (\varepsilon_h \cdot e_z \cdot e_z^* + h_z \cdot h_z^*) \cdot \frac{\varepsilon_h \cdot \omega^2}{q^2} \cdot \rho^2 \cdot \left( 1 + \frac{n^2}{q^2\rho^2} - \frac{K'^2_n(q\rho)}{K_n^2(q\rho)} \right)$$

$$- (\varepsilon_h \cdot e_z \cdot e_z^* + h_z \cdot h_z^*) \cdot \frac{\varepsilon_h \cdot \omega^2 + h^2}{q^3} \cdot \rho \cdot \frac{K'_n(q\rho)}{K_n(q\rho)}$$

$$W_{nk}^{(h)} = +\frac{L}{4} \cdot \frac{2i}{q^2} \cdot \frac{nh \cdot \varepsilon_h \omega}{q^2} \cdot (e_z^* \cdot h_z - e_z \cdot h_z^*) \cdot \gamma_{nk} \cdot \gamma_{nk}^* +$$

$$+ \frac{L}{4} \cdot (\varepsilon_h \cdot e_z \cdot e_z^* + h_z \cdot h_z^*) \cdot \frac{\varepsilon_h \cdot \omega^2}{q^2} \cdot \rho^2 \cdot \left( 1 + \frac{n^2}{q^2\rho^2} - \frac{K'^2_n(q\rho)}{K_n^2(q\rho)} \right) \cdot \gamma_{nk} \cdot \gamma_{nk}^*$$

$$- \frac{L}{4} \cdot (\varepsilon_h \cdot e_z \cdot e_z^* + h_z \cdot h_z^*) \cdot \frac{\varepsilon_h \cdot \omega^2 + h^2}{q^3} \cdot \rho \cdot \frac{K'_n(q\rho)}{K_n(q\rho)} \cdot \gamma_{nk} \cdot \gamma_{nk}^*$$

$$W_{nk}^{(h)} = +\frac{L}{4} \cdot \frac{2i}{q^2} \cdot \frac{nh \cdot \varepsilon_h \omega}{q^2} \cdot \left(e_z^* \cdot h_z - e_z \cdot h_z^*\right) \cdot \gamma_{nk} \cdot \gamma_{nk}^* +$$

$$+\frac{L}{4} \cdot \left(\varepsilon_h \cdot e_z \cdot e_z^* + h_z \cdot h_z^*\right) \cdot \frac{\varepsilon_h \cdot \omega^2}{q^2} \cdot \rho^2 \cdot \left(1 + \frac{n^2 - D_K^2}{q^2 \rho^2}\right) \cdot \gamma_{nk} \cdot \gamma_{nk}^*$$

$$-\frac{L}{4} \cdot \left(\varepsilon_h \cdot e_z \cdot e_z^* + h_z \cdot h_z^*\right) \cdot \frac{2\varepsilon_h \cdot \omega^2 + q^2}{q^4} \cdot D_K \cdot \gamma_{nk} \cdot \gamma_{nk}^*$$

$$W_{nk}^{(h)} = \frac{L}{4} \cdot \left(\varepsilon_h \cdot e_z \cdot e_z^* + h_z \cdot h_z^*\right) \cdot \frac{\varepsilon_h \cdot \omega^2 \cdot \rho^2}{q^2} \cdot \left(1 + \frac{n^2 - D_K^2 - 2D_K}{q^2 \rho^2}\right) \cdot \gamma_{nk} \cdot \gamma_{nk}^*$$

$$-\frac{L}{4} \cdot \left(\varepsilon_h \cdot e_z \cdot e_z^* + h_z \cdot h_z^*\right) \cdot \frac{D_K}{q^2} \cdot \gamma_{nk} \cdot \gamma_{nk}^* + \frac{L}{2} \cdot \left(e_z^* \cdot h_z - e_z \cdot h_z^*\right) \cdot \frac{inh \cdot \varepsilon_h \omega}{q^4} \cdot \gamma_{nk} \cdot \gamma_{nk}^*$$